\documentclass[twocolumn]{aastex631}
\usepackage{amsmath} 
\usepackage{CJK}      %%New add for chinese

\newcommand{\jz}[1]{{\color{black}{#1}}{}}
\newcommand{\jzn}[1]{{\color{black}{#1}}{}}
\newcommand{\jzl}[1]{{\color{black}{#1}}{}}  %

\newcommand{\fy}[1]{{\color{black}{#1}}{}}
\newcommand{\fyt}[1]{{\color{black}{#1}}{}}
\newcommand{\chg}[1]{{\color{black}{#1}}{}}

\newcommand{\chgt}[1]{{\color{black}{#1}}{}} %

\newcommand{\Hm}{H_{\rm m}}
\newcommand{\Ap}{\mathbf{A}_p}

\DeclareMathAlphabet{\mathbfit}{OML}{cmm}{b}{it}
\renewcommand{\mathbf}{\mathbfit}

\accepted{on 06-Apr-2023 by APJ} % A new received version, not the 2022-10-version
\shortauthors{Fu Yu et al.}   
\graphicspath{{./}{./}} 
\begin{document}
\begin{CJK*}{UTF8}{gbsn} %% New add for Chinese 

%% ===================================Title begin====================================
%% ===================================Title begin====================================
%% ===================================Title begin====================================
%\title{Comparison of Two Coronal Magnetic Field Models using MHS and CFITS methods in AR 12158\footnote{In prep on Dec, 16, 2021}}
%\title{Revisit the magnetic topology of AR12158 with force-field extrapolation}
%\title{The non-force-freeness matters for the extrapolation of solar coronal magnetic field?}
%\title{Does the non-force-freeness matter for the extrapolation of solar magnetic field?} constrained
\title{Magnetic field extrapolation in active region well \chg{comparable with} observations in multiple layers}

\correspondingauthor{Jie  Zhao}
\email{zhaojie@pmo.ac.cn}

\author[0000-0002-1713-2160]{Fu Yu}
\affiliation{Key Laboratory of Dark Matter and Space Astronomy, Purple Mountain Observatory, Chinese Academy of Sciences, Nanjing 210023, People's Republic of China}
\affiliation{School of Astronomy and Space Science, University of Science and Technology of China, Hefei 230026, People's Republic of China}
\affiliation{State Key Laboratory of Space Weather, National Space Science Center, Chinese Academy of Sciences, Beijing, 100190, People's Republic of China}

\author[0000-0003-3160-4379]{Jie  Zhao}
\affiliation{Key Laboratory of Dark Matter and Space Astronomy, Purple Mountain Observatory, Chinese Academy of Sciences, Nanjing 210023, People's Republic of China}
\affiliation{State Key Laboratory of Space Weather, National Space Science Center, Chinese Academy of Sciences, Beijing, 100190, People's Republic of China}

\author[0000-0002-4241-9921]{Yang Su}
\affiliation{Key Laboratory of Dark Matter and Space Astronomy, Purple Mountain Observatory, Chinese Academy of Sciences, Nanjing 210023, People's Republic of China}
\affiliation{School of Astronomy and Space Science, University of Science and Technology of China, Hefei 230026, People's Republic of China}

\author[0000-0002-1682-1714]{Xiaoshuai Zhu}
\affiliation{State Key Laboratory of Space Weather, National Space Science Center, Chinese Academy of Sciences, Beijing, 100190, People's Republic of China}

\author[0000-0002-9293-8439]{Yang Guo}
\affiliation{School of Astronomy and Space Science and Key Laboratory for Modern Astronomy and Astrophysics, Nanjing University, Nanjing 210023, People's Republic of China}

\author[0000-0003-4439-4972]{Jinhua Shen}
\affiliation{Xinjiang Astronomical Observatory, CAS, Urumqi, 830011, Peopleʼs Republic of China}

\author[0000-0003-1078-3021]{Hui Li}
\affiliation{Key Laboratory of Dark Matter and Space Astronomy, Purple Mountain Observatory, Chinese Academy of Sciences, Nanjing 210023, People's Republic of China}
\affiliation{School of Astronomy and Space Science, University of Science and Technology of China, Hefei 230026, People's Republic of China}

%\collaboration{6}{(AAS Journals Data Editors)}



\begin{abstract} %%%% ===========================================
% 2022 12 05 delete the old one
% 2022 12 22 new add from jz
%\jzl{
  Magnetic field extrapolation is a fundamental tool to reconstruct the three-dimensional magnetic field above the solar photosphere. However, the prevalently used force-free field model might not be applicable in the lower atmosphere with non-negligible plasma $\beta$, where the crucial process of flux rope formation and evolution could happen. In this work, we perform extrapolation in active region (AR) 12158, based on an recently developed magnetohydrostatic (MHS) method which takes plasma forces into account. 
By comparing the results with those from the force-free field extrapolation methods, we find that the overall properties, which are characterized by the magnetic free energy and helicity, \chg{are roughly the same}. The major differences lie in the magnetic configuration and the twist number of magnetic flux rope (MFR). Unlike previous works either obtained sheared arcades or one coherent flux rope, the MHS method derives two sets of MFR, which are highly twisted and slightly coupled. Specifically, the result in the present work is \chg{more comparable with} the high-resolution observations from the chromosphere, through the transition region to the corona, such as the filament fibrils, pre-eruptive braiding characteristics and the eruptive double-J shaped hot channel. Overall, our work shows that the newly developed MHS method is more promising to reproduce the magnetic fine structures that can well match the observations at multiple layers, and future data-driven simulation based on such extrapolation will benefit in understanding the critical and precise dynamics of flux rope before eruption.
%}
\end{abstract}

\keywords{
\href{http://astrothesaurus.org/uat/1518}{Solar photosphere (1518)}; 
\href{http://astrothesaurus.org/uat/1479}{Solar chromosphere (1479)};
\href{http://astrothesaurus.org/uat/1483}{Solar corona (1483)}; 
\href{http://astrothesaurus.org/uat/1493}{Solar extreme ultraviolet emission (1493)};
\href{http://astrothesaurus.org/uat/1974}{Solar active regions (1974)};
\href{http://astrothesaurus.org/uat/1503}{Solar magnetic fields (1503)}
}

%\keywords{Sun: magnetic fields---Sun: photosphere---Sun: chromosphere---Sun: corona}
%\keywords{Unified Astronomy Thesaurus concepts:}


\section{Introduction} \label{sec:intro}  %%===============================

%\begin{itemize}
%\item[•] Space weather: Flare, CME, magnetic field, etc
%\item[•] Method to get magnetic(advantage and dis$\sim$): MHS, NLFFF...
%\item[•] Why using MHS, or compare with NLFFF? Wiegelmann, Zhu, Zhao
%\item[•] Our work: other done in AR12158 but ...; also less case done comparing MHS and NLFFF.
%\item[•] Just demo here ......
%\end{itemize}

Solar eruptions, especially those from active regions (ARs), are acknowledged to be the major driving sources of catastrophic space weather. The accompanying coronal mass ejections (CMEs) have become an important target for space weather monitored in remote observations and in in-situ measurements \citep[][]{Zhang_jie2021}. It is generally believed that the physical essence of CMEs is the eruption of magnetic flux ropes \citep[MFRs;][]{Chen_james1996, Titov1999} that carrying plasma  \citep[][]{Chen_peng_fei_2011_cme}. Therefore, studying the structure and evolution of MFRs is crucial to understand the physical mechanisms of CMEs and predict space weather. As it is currently difficult to measure the three-dimensional (3D) coronal magnetic field directly from observations, an alternative way of modeling coronal field with \chg{observed photospheric field used as boundary condition \citep{Sakurai1989, Solanki2006, Wiegelmann2017, Wiegelmann2021, Zhu_XS_2022_review} or with the rarely obtained chromospheric field \citep{Wiegelmann2008_use_chro, Harvey2012, Jin2013, Lagg2017} has been implemented. 
}

% 0811 before delete

Overall, magnetohydrodynamics (MHD) approximation can provide a basic framework to outline the interaction between plasma and magnetic field in the solar atmosphere \citep{Priest2014}. It is simplified under various conditions according to the two critical parameters of plasma $\beta$ and $\rm{Alfv\acute{e}n}$ Mach number $M_{\rm{A}}$ \citep{Gary2001, Wiegelmann2017}. Plasma $\beta$ evaluates the dominance of the plasma gas pressure over the magnetic pressure, while the $\rm{Alfv\acute{e}n}$ Mach number estimates the flow speed over the $\rm{Alfv\acute{e}n}$ speed. Generally, MHD approximation should be adopted for the outer corona where the plasma flow and plasma gas pressure dominate. In regions with low $M_{\rm{A}}$ and finite $\beta$, e.g., in the upper photosphere and chromosphere, a steady-state derives
% 2022 10 07 change
%Generally, MHD approximation should be adopted for the outer corona where the plasma flow and plasma gas pressure dominate, while MHS approximation could be considered for the inner corona where slow dynamic processes are treated as being quasi-static. For the latter situation, the steady-state derives
% 2022 08 11 change is as It will ...
%Considering that some slow dynamic processes of the corona are quasi-static, i.e., \chg{typically} with $\rm{Alfv\acute{e}n}$ Mach number (fluid speed over $\rm{Alfv\acute{e}n}$ speed) $ M_{\rm{A}} \ll 1 $, the steady-state approximation derives
the magnetohydrostatic (MHS) model with the Lorentz force being balanced by the plasma pressure gradient and gravitational forces \citep{Wiegelmann2017}. Under a further simplified condition of low $M_{\rm{A}}$ and low $\beta$, e.g., in the inner corona, the Lorentz force vanishes and the magnetic field becomes force-free, which is a state that  assuming the dominance of the magnetic field over plasma. Models including nonlinear force-free field (NLFFF), linear force-free field (LFFF) and potential field are often applied to different force-free states (or current distributions) in terms of the force-free parameter \citep[noted as $\alpha$;][]{Wiegelmann2008, Wiegelmann2021}.
In particular, NLFFF extrapolations are favored for its relatively robust reconstructions and affordable computational resources in most cases. Various numerical methods have been developed to construct the NLFFF models, e.g., the optimization approach \citep{Wheatland2000, Wiegelmann2004}, magneto-frictional method \citep{Valori2007, Guo_yang2016}, MHD relaxation method \citep{Jiang_chaowei2012}, Grad-Rubin method \citep{Grad_1958, Amari2006, Gilchrist2014_CFITS} and flux rope insertion method \citep{van_Ballegooijen2004, Su_ying_na2019}.
%test 2022 10 07
%{\color{red}In the inner corona (low beta, low Alfven Mach number) a force-free approach is justified. MHS should be done in regions with low Alfven Mach number and finite beta, e.g. in the upper photosphere and chromosphere.}

%% YF 2022 08 05 New add delete

Based on the force-free field methods, investigations have been carried out for understanding the initiation mechanism and the process of energy release during solar eruptions. Most of the works obtain an eruptive MFR with twist number around 1, and/or \jzn{with} decay index above the eruptive flux rope around 0.5 $\sim$ 2 \citep{Sun_XD2022} or 0.8 $\sim$ 1.5 \citep{Zhong2021NC}, which are consistent with the predictions of the kink instability and/or the torus instability. 
\jzl{
Nevertheless, the interplanetary magnetic clouds (MCs) originating from the sun however are found to have a wide range of twist number as high as 14.6 turns per astronomical unit from in-situ observations \citep{Hu_qiang2014}.
%while some work found the free energy and the magnetic reconnection responsible for the trigger locate at the low atmosphere (Zhao 2014).
  Besides the possibility that the flux rope may evolve during its propagation in the interplanetary space \citep{Wang_YM2018}, the discrepancy of the twist number obtained near the sun and in-situ may also lie in the force-free assumption that has been adopted for the \chg{entire atmosphere}.} By the MHD relaxation method, \jzl{it is found that the non-force-free region can reach height of 1.4 $\sim$ 1.8 Mm above the photosphere \citep{Zhu_XS_2016} and the non-force-freeness in the chromosphere has also been revealed in radiative MHD simulation \citep{Leenaarts2015}}. 

% note 2022 12 05  

For attaining a reliable picture of the solar eruption, it is essential to study the magnetic configuration under non-force-free assumption that matches better with the physical situation, especially in the lower atmosphere. \jzl{Therefore, an MHS extrapolation method \citep{Zhu_XS_2018} is developed for computing such magnetic configuration by taking into account the plasma pressure and gravity.}
%force-field assumption based on magnetohydrostatic equilibria (as mentioned above).
%Based on the radiative MHD simulation \citep{Cheung2019}, comparison was carried out for the MHS and the NLFFF extrapolation \citep{Zhu_XS_2019}. 
It is found that, by applying both MHS and NLFFF extrapolation to a radiative MHD simulation \citep{Cheung2019}, the MHS method has more advantages in reconstructing long twisted magnetic field lines and in recovering the primary plasma structure in the lower atmosphere \citep{Zhu_XS_2019}.

%note 2022 12 05， about MHS in low-atmosphere
%With observation of 

\jzl{
%As the observation is the most important benchmark for validating any reconstruction, it is often compared in various works. 
Comparison between observation and the extrapolation result is essential as the observation is the most important benchmark for validating any reconstruction method.
For the NLFFF extrapolations, the associated QSLs could well define the main features during eruption, such as the initial position of flare ribbons and also the hot channels \citep{2006SoPh..238..293M,2011SoPh..269...83C,Zhao_jie2016,2016ApJ...817...43S}. For the MHS method, the field lines often reflect the chromospheric fine structures, such as the fibrils \citep{Zhu_XS_2016, Zhu_XS_2020} and brightenings \citep{Zhao_jie2017}. 
%The magnetic field obtained from MHS methods also show similarity with the chromospheric magnetic field when it is achieved \citep{Zhu_XS_2020,Vissers2021}.
It turns out that different methods have their advantages in different aspects and it is necessary to make an integration.
%By combining force-free and non-force-free extrapolation at different layers, an updated MHS method is developed \citep{Zhu_XS_2022}.
}

\jzl{In this paper, we aim to show the performance of the MHS extrapolation
%, which combines force-free assumption at upper layers and non force-free assumption at lower layers, 
in reconstructing 3D magnetic field of an active region. For this intention, AR 12158 which has been extensively studied \citep{ Cheng_xin2015, Zhao_jie2016, Zhou_guiping2016, Vemareddy2016, Duan2017,Lee2018, Kilpua2021,shen_jinhua2022} is selected and we carry out the extrapolation with newly updated MHS method \citep{Zhu_XS_2022}. The results are analyzed in details and are 
compared with the ones from other works.
The paper is organized as follows: the observation characteristics are shown in Section \ref{subsec:obs} and the extrapolation method is introduced briefly in Section \ref{subsec:model_mhs}. The results are displayed in Section \ref{sec:res} and we present the summary and discussions in Section \ref{sec:discussion}.}

\section{Observations and Extrapolation Methods}\label{sec:obs_model}%%===============================
%\section{Extrapolation Methods}\label{subsec:model_intro}
\subsection{Observations}\label{subsec:obs}

%% The "ht!" tells LaTeX to put the figure "here" first, at the "top" next
%% and to override the normal way of calculating a float position
\begin{figure*}[ht!]
%\plotone{obs_views.pdf}    %%====================================
%\plotone{obs_views_v5_.pdf}  %%===================================
%\plotone{mag3d_braid_v5.pdf}    %%================================
%\plotone{obsv05.pdf}   
\plotone{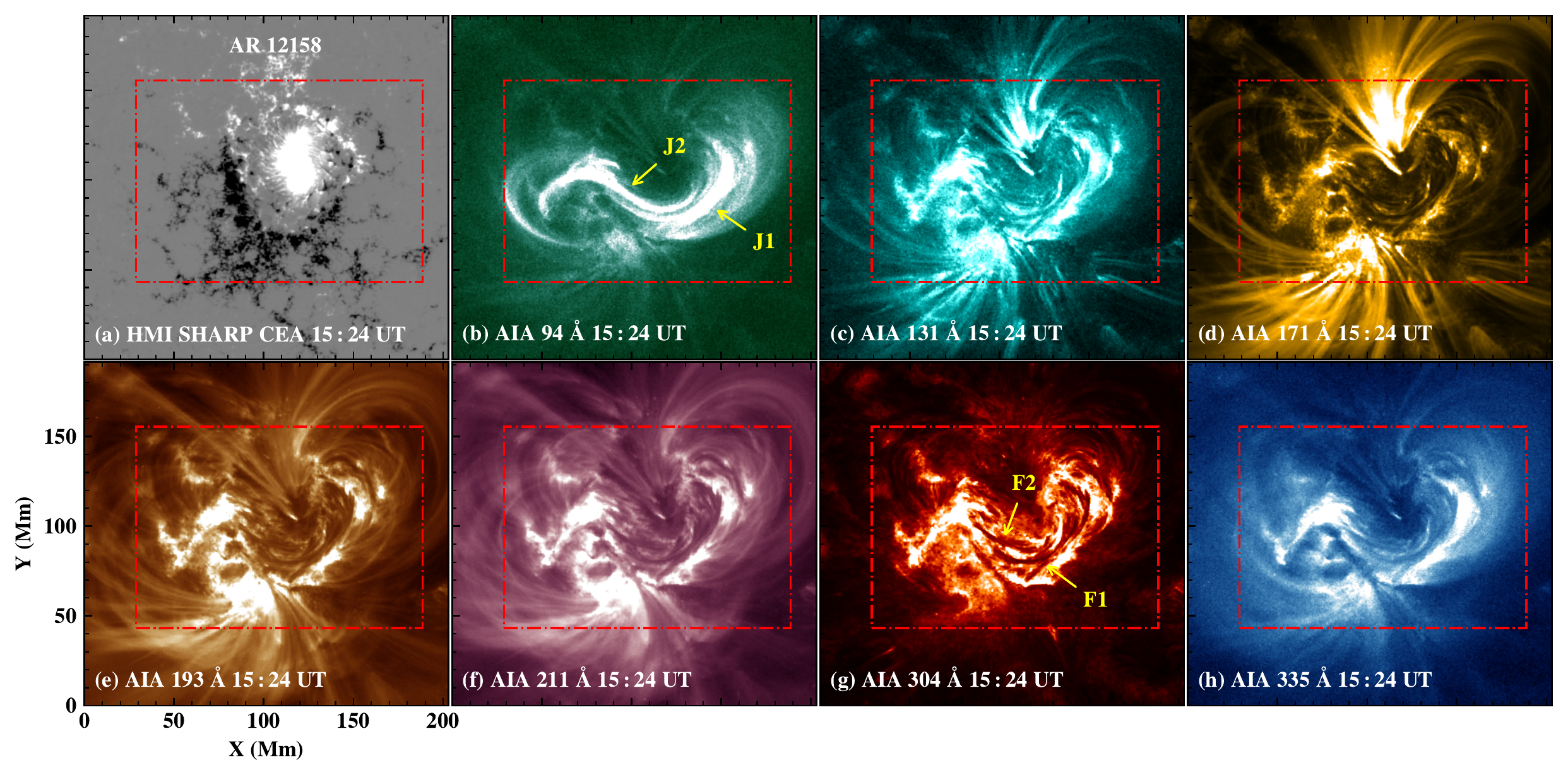}    %%=====================================

\caption{\fy{SDO observations of AR 12158 at 15:24 UT on 2014 September 10. Panel (a) shows the SHARP CEA remapped line-of-sight (LOS) magnetogram from SDO/HMI. Panels (b)$-$(h) show images of seven EUV channels from SDO/AIA. All the images are coaligned, and the red dashed boxes denote the major domain for the analyses in Section \ref{sec:res}. 
An associated animation of panels (b)$-$(h) is available. The animation shows the evolution of various features such as sigmoid, hot channel, flare ribbon and the possible magnetic braiding during the period from 12:00 UT to 17:20 UT. The animation duration is about 64 s.\\
(An animation related to this figure is available.) }
\label{fig:obs_view}}
\end{figure*}
%% SDO observations of AR12158 at the time 15:24 UT on 2014 September 10.
% before 08 17
%  NOAA AR 12158, which is selected for ...

% ==== Obs: AIA、HMI
  NOAA AR 12158, which is selected for this case study, is observed near the disk center (N15E03 in heliographic coordinates) on 2014 September 10 by the Solar Dynamics Observatory \citep[SDO;][]{sdo2012}. During its passage through the solar disk, an X1.6 class flare happens at 17:21 UT according to the soft X-ray 1--8 $\rm{\AA}$ flux from the Geostationary Operational Environmental Satellites (GOES), and a halo coronal mass ejection (CME) event is subsequently accompanied (see CME catalog\footnote{\url{https://cdaw.gsfc.nasa.gov/CME_list/}} \citep{Gopalswamy2009} from Large Angle and Spectrometric Coronagraph \citep[LASCO;][]{Brueckner1995} aboard the Solar and Heliospheric Observatory \citep[SOHO;][]{Domingo1995}).
  %, implying the ejection of MFR as well as plasma \citep{Chen_peng_fei_2011_cme}. 
  The observations from SDO at pre-eruptive phase are shown in Figure \ref{fig:obs_view}.

  \jz{The magnetic field is observed by the Helioseismic and Magnetic Imager \citep[HMI;][]{sdo_hmi2012} and the product of Spaceweather HMI Active Region Patch (SHARP) series \citep{sdo_hmi_sharp2014, sdo_hmi_2014} is usually \chg{employed for} magnetic field extrapolation. For data labeled with '$\rm{hmi.sharp\_cea\_720s}$' as displayed in panel (a), it performs the projection and remapping of the observed vector magnetic field onto the heliographic Cylindrical Equal-Area (CEA) coordinate system \citep{Sun2013_hmi}, and an equal pixel size (0.36 Mm) is derived in both longitude (or X) and latitude (or Y) direction.}
  
  \jz{Images of seven EUV channels from Atmospheric Imaging Assembly \citep[AIA;][]{sdo_aia2012} are shown in panels (b)--(h). \chg{An overall bright sigmoid structure is clearly seen in AIA 94 $\rm{\AA}$ even at the pre-eruptive phase.} Specifically, \chg{the overall sigmoid in AIA 94 $\rm{\AA}$ shows separation in the middle, and the two separated parts which are pointed by arrows with J1 and J2 are the so-called double J-shaped structure. Filament fibrils, such as the ones pointed by arrows F1 and F2 in panel (g) are found in other wavelengths.}
%the overall sigmoid in AIA 94 $\rm{\AA}$ seems to consist of two separated parts \chg{(pointed by arrows with J1 and J2)}, and filament fibrils \chg{(such as F1 and F2 in panel (g))} are found in other wavelengths. 
  The overall sigmoid structure has been reproduced with various extrapolations \fy{\citep{Zhao_jie2016, Kilpua2021}} while the fine structures, e.g., the dark fibrils, have no apparent correspondences in the \jzl{previously} obtained 3D magnetic field.} 
%, while double J-shaped feature is also identified in other EUV wavelengths
%\chgt{For further investigation, we will also check the high-resolution observations from the Big Bear Solar Observatory (BBSO)/Goode Solar Telescope (GST) in  Near Infra-Red channel, i.e. He I 10830 $\rm{\AA}$ \citep{Cao2012_bbso} and the Interface Region Imaging Spectrograph \citep[IRIS, ][]{DePontieu2014_iris}/slit-jaw imager (SJI) in 1400 $\rm{\AA}$ and 2796 $\rm{\AA}$,  and compare them with the MHS extrapolation results in Section \ref{subsec:mag_brt}. }

\chg{To better understand the detailed features mentioned in Figure \ref{fig:obs_view}, the EUV images at five selected time steps with a cadence of 1 hour are displayed in the top panels of Figure \ref{fig:obs_and_mag} from left to right. 
%The field lines from MHS model are shown in the bottom panels (e)--(g), with different field line systems overlaying on different features observed in AIA $\rm 131 \AA$, $\rm 304 \AA$, $\rm 94 \AA$ from left to right.
The plus signs are added and annotated with 'A'--'E' for identifying the locations of different features. From 12:13 UT to 14:11 UT, the bright structure between B and C (pointed by the cyan arrows) evolves gradually. 
% YF 0217
% Enhanced emission (pointed by white arrows) is found in AIA 94 and 131 $\rm \AA$ at 14:11 UT through a winding influence at 13:19 UT. 
%\chg{This phenomenon and evolution characteristics suggest the existence of magnetic rope flux with large twist, as well as the possible evidence of braiding and magnetic reconnection \citep{Awasthi2018}.}
%The winding is also identified in AIA 171 and 304 $\rm \AA$ while the induced emission enhancement is relatively weak. 
%
Such bright structure includes fibril-like features, which show overcrossing morphology at 12:13 UT and even twist at 13:19 UT. Apparent enhanced emission (pointed by white arrows) is found in AIA 94 and 131 $\rm\AA$ at 14:11 UT. 
The above phenomenon and evolution characteristics may suggest the existence of magnetic flux rope with large twist, which is reproduced in our extrapolation that is introduced in Section 3.1, as well as the possible evidence of braiding and magnetic reconnection \citep{Awasthi2018}. The twist fibril-like feature is also identified in AIA 171 and 304 $\rm \AA$ at 13:19 UT while the relevant emission enhancement at 14:11 UT is relatively weak.
The sigmoid structure starts to brighten at 15:24 UT in AIA 94 $\rm{\AA}$, and a slender thread of brightening (pointed by yellow arrows) becomes visible in AIA 131, 171, 304 $\rm{\AA}$ at 16:34 UT. These evolutionary processes also reveal the independence of J1 and J2 as separate parts and their relevance as a overall sigmoid.
\jz{The observed} sophisticated evolution before the eruption may imply the existence of complex magnetic structure in the active region, such as the braiding magnetic field lines \citep{Parker1983a, Berger2009, Cirtain2013Nature, Pontin2017} and the multi-flux-rope system \citep{Awasthi2018}. A detailed comparison with the magnetic field is shown in Section \ref{subsec:mag_brt}.
}
  
  \jzl{High resolution observations are also available for this event. The lower chromosphere is observed by the Big Bear Solar Observatory (BBSO)/Goode Solar Telescope (GST) in  Near Infra-Red (NIR) channel, i.e., \ion{He}{1} 10830 $\rm{\AA}$  with a pixel size of 0.$\rm \arcsec$079 \citep{Cao2012_bbso}. The slit-jaw images (SJI) in Ultra-Violet (UV) wavelengths, i.e., 2796 $\rm{\AA}$ for the chromosphere and 1400 $\rm{\AA}$ for the transition region, are observed by the Interface Region Imaging Spectrograph \citep[IRIS,][]{DePontieu2014_iris} with a pixel size of 0.$\rm \arcsec$16. \chg{The observations with partial field of view (FOV) are displayed in Figure \ref{fig:obs_high_res} for further comparison in Section \ref{subsec:mag_brt}. 
}
}

%  Near Infra-Red Imaging Spectropolarimeter
  %whether it is the high-temperature bright structures in AIA 94 $\rm{\AA}$, the low-temperature dark filaments in AIA 304 $\rm{\AA}$, and the similar features in other passbands. 

% ==== Obs: HMI
%  On the other hand, the vector magnetograms from the Helioseismic and Magnetic Imager \citep[HMI;][]{sdo_hmi2012} on board SDO, specifically the Spaceweather HMI Active Region Patch (SHARP) version is applied to magnetic field extrapolation which can be accessed through Joint Science Operations Center (JSOC). In addition, SHARP data series \citep{sdo_hmi_sharp2014, sdo_hmi_2014} particularly have two versions, namely, one is $\rm{hmi.sharp\_720s}$ which directly cutout from the full-disk images in the Helioprojective-cartesian CCD coordinate, and the other is $\rm{hmi.sharp\_cea\_720s}$ which performed the projection and remapping \citep{Sun2013_hmi} onto the heliographic Cylindrical Equal-Area (CEA) coordinate system thus result in equally spaced pixels (0.36 Mm per pixel) in both longitude(or X) and latitude(or Y) such as shown in Figure \ref{fig:obs_view}(a). 

%\section{Extrapolation Methods}\label{subsec:model_intro}
\subsection{MHS extrapolation method}\label{subsec:model_mhs}
%, which combines force-free assumption at upper layers and non force-free assumption at lower layers, 

  The MHS model which solves the MHS equations by optimization method is developed by \citet{Zhu_XS_2018} and optimized by \citet{Zhu_XS_2019}. To improve the computational efficiency, \citet{Zhu_XS_2022} combines the MHS and NLFFF methods in which the MHS method is constrained in the very lower layer to produce magnetogram at a nearly force-free height for the NLFFF extrapolating upward. 
  The latest version of the model \citep{Zhu_XS_2022} is applied in the present work to obtain 3D magnetic field. The MHS equilibrium is described by the following equations:
%, which considers MHS state in the lower atmosphere and force-free field in the upper atmosphere, 
\begin{equation}  
  \frac{1}{\mu_0}(\nabla \times \mathbf{B}) \times \mathbf{B} - \nabla p -\rho g \hat{z} = 0,\ \rm{satisfying} \ \nabla \cdot \mathbf{B} = 0. \label{eq:model}
\end{equation}
where $\mathbf{B}$, $p$, $\rho$, are the vector magnetic field, plasma pressure and plasma density, while $g$ and $\mu_0$ denote the gravitational acceleration and vacuum permeability, respectively. More details can be found in \citet{Zhu_XS_2022}. 

%\chgt{
%While under the assumption of force-free field, the second and third terms in the MHS equation (Equation (\ref{eq:model})) vanish and only the term of magnetic force remains, namely $\mathbf{J} \times \mathbf{B} = 0$.}

To reproduce the 3D magnetic field of AR 12158 with the MHS model, the vector magnetogram from SHARP CEA is adopted as the photospheric boundary as the MHS model is established in the Cartesian coordinate system. 
The radial component \jz{($B_r$)} of the \jz{photospheric} magnetic field, \chg{which has original size of $\rm 564 \times 532$ $\rm pixel^{2}$, is shown in Figure \ref{fig:obs_view}(a). A volume of $\rm 444 \times 312 \times 200$ $\rm pixel^{3}$ with uniform pixel size of 0.36 Mm is extracted from the extrapolation result, and the bottom boundary is shown by the dashed box in Figure \ref{fig:obs_view}(a).}

%A volume with uniform pixel size of 0.36 Mm and size of $\rm 444 \times 312 \times 200$ $\rm pixel^{3}$

% 2022 12 05 delete CFITS 

\section{Results} \label{sec:res} %===========================
%\subsection{Magnetic field and the pre-eruption brightenings}\label{subsec:mag_brt}
\subsection{Magnetic field and the pre-eruption \jzl{fine structures}}\label{subsec:mag_brt}

%% The "ht!" tells LaTeX to put the figure "here" first, at the "top" next
%% and to override the normal way of calculating a float position
\begin{figure*}[ht!]
\centering
\plotone{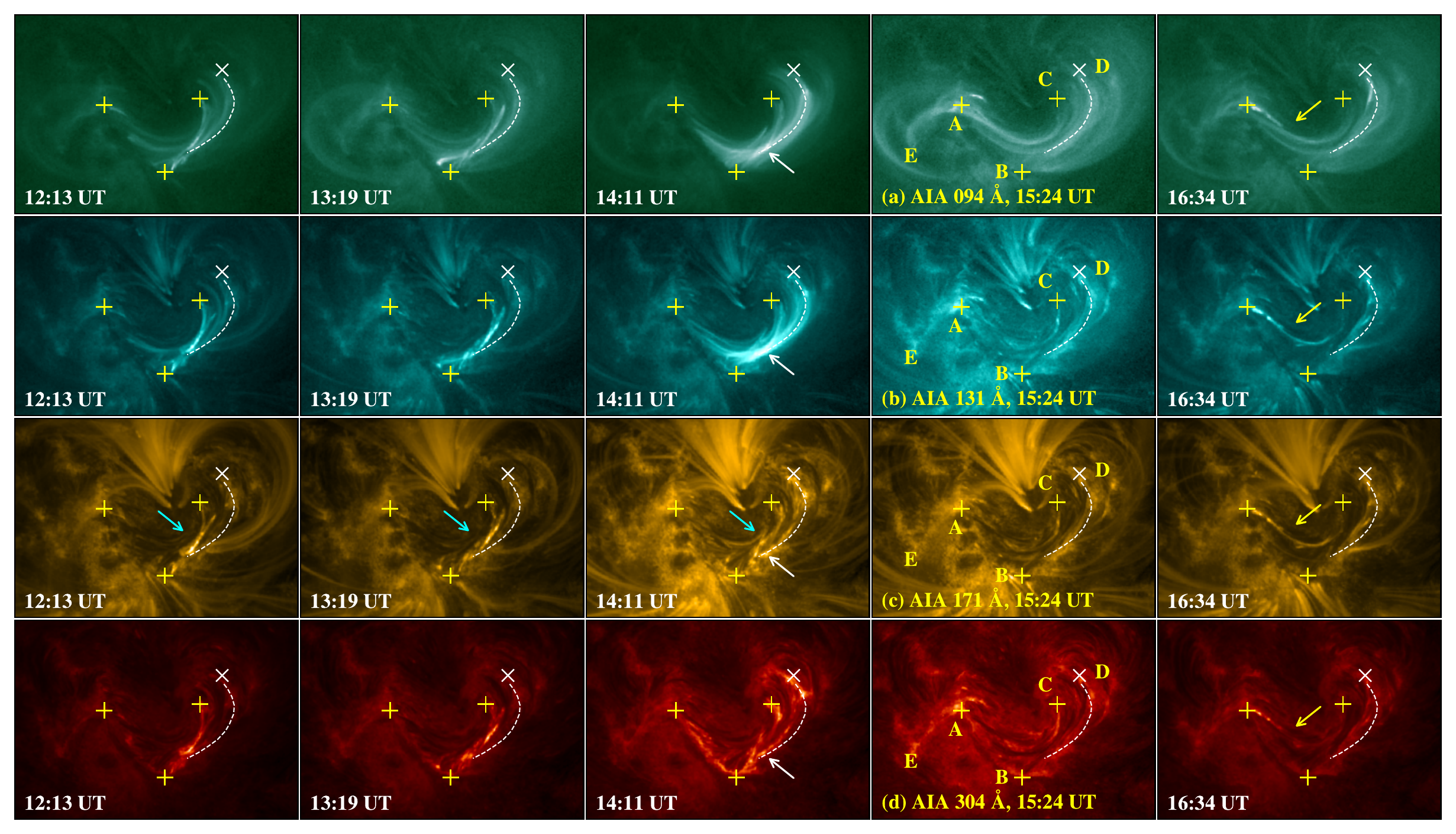}  %%==================================
%\plotone{mag3d_braid_v6.pdf}  %%=======================================
\centering
\plotone{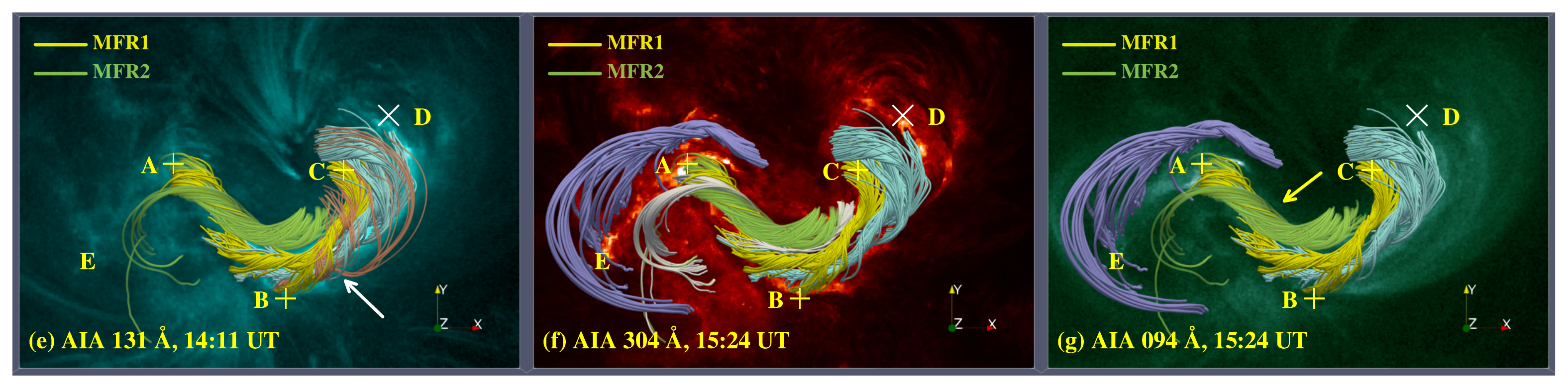}  %%=======================================
%\centering
%\plotone{high_obs02__.pdf}  %%=======================================

% 2022 12 20 note
%\caption{SDO observations at multiple time steps and the reconstructed magnetic field. Top panels: time evolution of the event in EUV images in 94, 131, 171, and 304 $\rm{\AA}$ at five time steps (with 1 hour cadence) before the X-class flare. Bottom panels: the reconstructed magnetic field from MHS at 15:24 UT are overlaid on the AIA images at 14:11 UT for 131 $\rm{\AA}$, 15:24 UT for 304 and 94 $\rm{\AA}$ for comparison. Different colors show the magnetic field lines with different connectivity. The plus signs and dash lines in the top and bottom parts are labeled for the reference of the locations for different structures. 
%\label{fig:obs_and_mag}}
\caption{SDO observations at multiple time steps and the reconstructed magnetic field. 
Top panels: time evolution of the event in EUV images from SDO/AIA 94, 131, 171, and 304 $\rm{\AA}$ at five time steps (with 1 hour cadence) before the X-class flare. Bottom panels: the reconstructed magnetic field lines from MHS extrapolation at 15:24 UT are overlaid on the AIA images at 14:11 UT for 131 $\rm{\AA}$, 15:24 UT for 304 and 94 $\rm{\AA}$ for comparison. Different colors show the magnetic field lines with different connectivity. 
 All the plus signs and dash lines in the panels are labeled for the reference of the locations for different structures. 
\label{fig:obs_and_mag}}
\end{figure*}

\begin{figure*}[t!]
\centering
\plotone{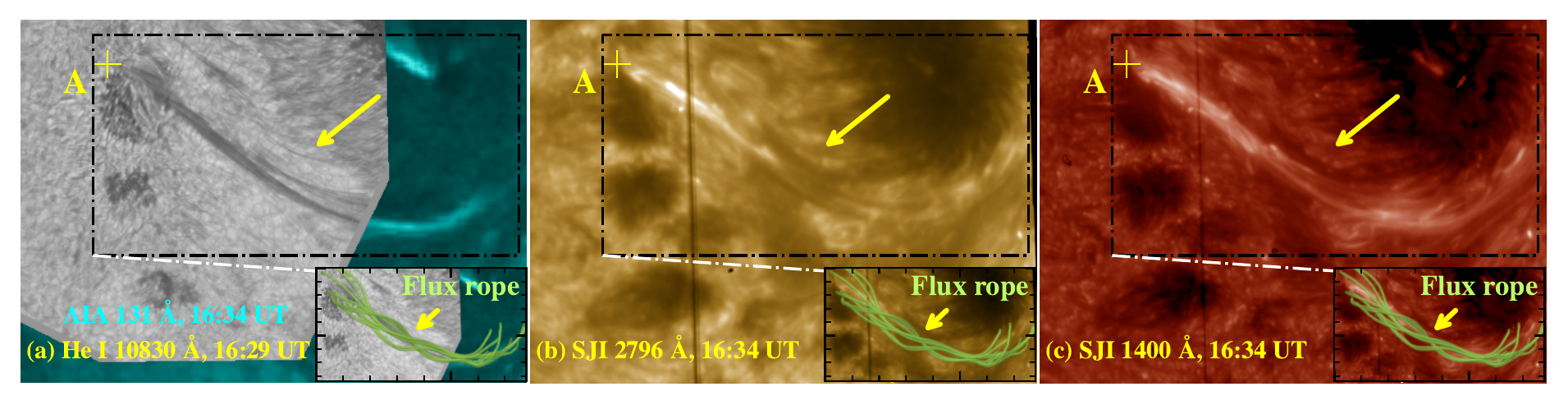}  %%========================================
\caption{High-resolution \jzl{observations} with \jzl{partial} field of view (FOV) as an extension of Figure \ref{fig:obs_and_mag}. Panel (a): the grayscale image is \jzl{\ion{He}{1}} 10830 $\rm{\AA}$ taken by BBSO/GST at 16:29 UT, and its missing FOV is filled with AIA 131 $\rm{\AA}$ at 16:34 UT as the background. Panels (b)--(c) are observations from SJI 2796 and 1400 $\rm{\AA}$ at 16:34 UT. 
\jzl{The box at the lower-right corner of each panel shows the top view of magnetic flux rope overlaid on the corresponding sub-region.} 
\label{fig:obs_high_res}}
\end{figure*}

%	To further analyze the compatibility of the MHS extrapolation, and to verify what was considered in Section \ref{sec:obs} of Figure \ref{fig:obs_view}, 
The brightenings of the sigmoid structure before eruption have been identified in various observations \fy{\citep{McKenzie2008, Zhang_Jie2012, Vemareddy2014, Cheng_xin2014, Cheng_xin2016}} and also in the present work (see movie of Figure \ref{fig:obs_view}). They are suspected to be associated with magnetic reconnection, which is crucial for understanding the initiation and eruption of the flux rope system. The MHS model has an advantage in constructing the low-lying magnetic field lines, which enables a better comparison between the pre-phase reconnection and the magnetic field in the low atmosphere. The reconstructed magnetic field from MHS model is analyzed in detail to show its correspondence \jzl{firstly} with the brightenings in AIA observations \jzl{which covers a range from chromosphere to the corona, and then specifically with the dark/bright fibrils in the extremely high-resolution observations at different layers of chromosphere and the transition region. }
\chg{The field lines from MHS model are displayed in the bottom panels (e)--(g) of Figure \ref{fig:obs_and_mag}, with color-coded field-lines' groups overlaying on different features observed in AIA $\rm 131 \AA$, $\rm 304 \AA$, $\rm 94 \AA$ from left to right. In general, there are two main flux ropes represented by the field lines in yellow (MFR1) and green (MFR2) respectively. The surrounding field lines intimately related to MFR1 are coded with cyan and orange, while the left loop system is shown in purple.}
%The \jzn{relevant} field lines \jzn{are displayed in the bottom panels, with different connectivities shown with different colors} for a better demonstration. 
%In general, there are two main flux ropes represented by the field lines in yellow (MFR1) and green (MFR2) respectively.
% 08 11 change
%In general, there are two main flux ropes (MFR1 and MFR2 which are determined from quasi-separatrix layers (QSLs) as shown in Section \ref{subsec:qsl_brt}), represented by the field lines in yellow and green respectively. 
%Ambient field lines that intimately coupled with the flux rope are displayed in purple, cyan and orange.
%Two main flux ropes, which are determined from quasi-separatrix layers (QSLs) as shown in Section \ref{subsec:qsl_brt}, and the ambient field lines are displayed.

%Meanwhile, the MFR2 between the dot-dashed lines b and c in Figure \ref{fig:qsl_2d} (f) and the bright structures between b' and c' in panels (a) and (b) also confirm their correspondence.

The field lines in cyan\jzn{, which} are beneath MFR1 for the left part and over-wrapping the latter at the right end\jzn{,} correspond to the dark region between C and D in AIA 304 $\rm{\AA}$ (see \jzn{panel }(f)). The combination of MFR1 and field lines in cyan may represent the emission enhancement of AIA 94, 131$\rm{\AA}$ at 14:11 UT. The region indicated by the white arrow \jzn{in panel (e)}, which has \jzn{over-wrapping field lines in orange on top of the MFR1 and field lines in cyan, may be responsible for the braiding features before 14:11 UT.}
%both the twisted MFR and the braided field lines, matches the brightest site well.
\chg{
In addition, the right branch of MFR1 between B and C seems to be consistent with the other dark stripe that is next to the aforementioned predominant one between C and D in AIA 304 $\rm{\AA}$. The field lines in purple have their feet anchored near the bright belt between A and E of AIA 304 $\rm{\AA}$, while the loops themselves  tend to coincide with the outer loops of AIA 94 $\rm{\AA}$ as shown in panel (g). 
}

The primary part of MFR2 (field lines in green) is cospatial with the slender emission between A and B at 16:34 UT, and the left extension tends to outline the loop between A and E of AIA 94 $\rm{\AA}$. 
\chgt{In particular, in the high-resolution images \jzl{of \ion{He}{1} 10830 $\rm{\AA}$, SJI 2796 and 1400 $\rm{\AA}$ as shown in Figure \ref{fig:obs_high_res} }, the slender dark/bright fibrils are in good agreement with the position of the magnetic flux rope MFR2 \jzl{as shown in the lower-right corners}. In Figure \ref{fig:obs_high_res}(a), the \jzl{\ion{He}{1}} 10830 $\rm{\AA}$ image has \jzl{a smaller} FOV, and its missing \jzl{part} is filled with AIA 131 $\rm{\AA}$. We \jzl{notice that the cold component of the slender feature in 10830 $\rm{\AA}$ is well connected to the hot component in 131 $\rm{\AA}$, indicating the multi-temperature components for such structure.}}
%\jzl{???} These imply }the early stage of the hot channel suggested by \citet{Cheng_xin2015} and the pre-existing MFR by \citet{Zhou_guiping2016}. 

%\jzn{
%In addition, the right branch of MFR1 between B and C seems to be consistent with the other dark stripe that is next to the aforementioned predominant one between C and D in AIA 304 $\rm{\AA}$. The field lines in purple have their feet anchored near the bright belt between A and E of AIA 304 $\rm{\AA}$, while the loops themselves  tend to coincide with the outer loops of AIA 94 $\rm{\AA}$ as shown in panel (g). 
%} move to before
%The dark stripe passed by the dot-dashed line b' in Figure \ref{fig:qsl_2d} (a) implies the magnetic structure coded by white color in panel (f).
%\chgt{add high resolu}

\chgt{
The \jzl{above} comparison between magnetic field lines of MHS extrapolation and observations from \jzl{multiple layers, ranging from chromosphere consectively to the corona, reveals} good correspondence, which is rarely achieved. 
%\chg{It suggests the effectiveness of MHS assumption in recovering the fine structures of sigmoid system in the present work.}
}

%delete old notes

\subsection{EUV Eruptive Features and QSLs}\label{subsec:qsl_brt}
 \label{subsec:qsl}%%===============================================================

\begin{figure*}[ht!]
%\plotone{logQ_cfits_mhs_v7.pdf}%%==================================
%\plotone{logQ_cfits_mhs_v9-127.pdf}%%================================
%\plotone{logQ_mhs_v1_.pdf}%%================================
\plotone{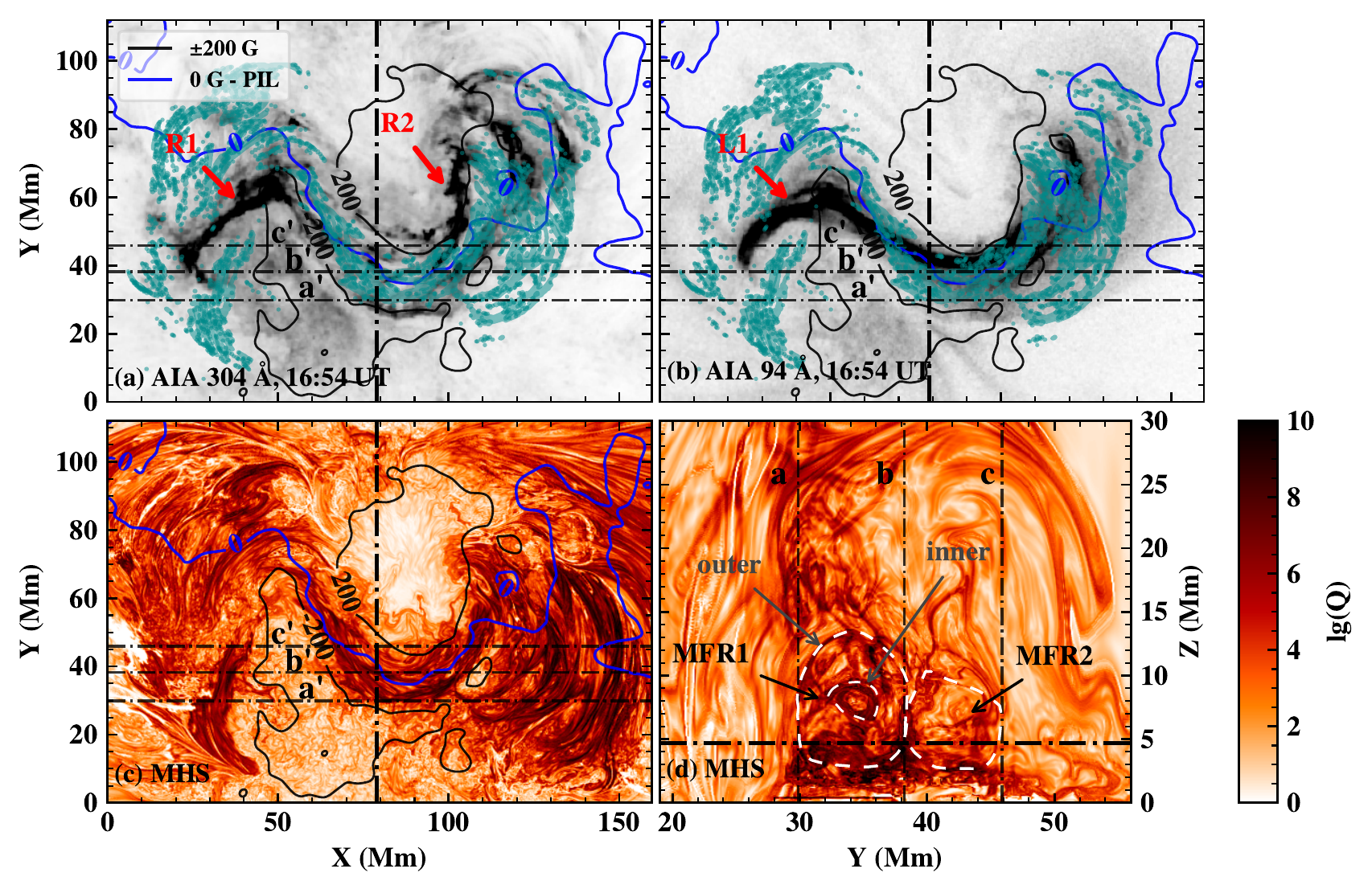}%%================================
%\plotone{mag3d_braid_v2.pdf}    %%==================================

\caption{Comparison between the eruptive features and the calculated squashing factor $Q$ of the magnetic field.
\chg{The AIA 304 and 94 $\rm{\AA}$ images with reversed grayscale} are shown in panels (a) and (b), \chg{respectively. Brightenings are displayed in dark color.} The 2D distributions of \chg{$Q$} are displayed in panels (c) and (d). \chg{The contours of QSLs from panel (c) with lg($Q$) = 6.5 are overlaid on panels (a) and (b) in green, by extracting the relevant parts corresponding to the brightening regions in AIA 304 $\rm{\AA}$}. The blue and dark contours in panels (a)--(c) show the $B_z$ component from the potential field, with levels of 0 and ±200 G. The vertical dot-dashed lines in panels (a), (b) and (c) denote the slice position of panel (d) while the horizontal dot-dashed line in panel (d) shows the \jzl{layer} where the $Q$ \chg{values} are sampled and displayed in panel (c).
%The dot-dashed lines annotated with a, b and c in panel (d) outline the region of MFR1 and MFR2 in the direction of Y-axis. 
\chg{Three vertical planes are used to identify the location of the flux ropes, which have intersections on the lines a, b, c on the vertical cut shown in panel (d), and a', b', c' on the horizontal cut in panel (c). The white dashed lines in panel (d) are manually selected to mark the boundary of the flux ropes.
}
% delete CFITS
 \label{fig:qsl_2d}}
\end{figure*}

%The correspondence between the observations and the magnetic field of the MHS model in the aspect of pre-eruption brightening is already demonstrated in Section \ref{subsec:mag_brt}. 
To show the ability of the MHS modeling in inferring the \jz{eruptive features}, such as flare ribbons and hot channels, the squashing factor $Q$ are calculated \jzl{and the results are displayed in Figure \ref{fig:qsl_2d}.}

The $Q$ factor measures the gradients of magnetic connectivity in a volume, and the region with large $Q$ value is defined as quasi-separatrix layer \citep[QSL,][]{Demoulin1996, Titov2002}, which separates the region of distinct flux systems. Hence, $Q$ map is a good indicator for ascertaining the characteristic structure of magnetic field. The method for calculating the $Q$ factor can be found in \citet{Zhao_jie2014} and can also refer to a FORTRAN parallel routine from \citet{Liu_R2016}\footnote{\url{http://staff.ustc.edu.cn/~rliu/qfactor.html} or an updated site \url{https://github.com/el2718/FastQSL} according to \citet{Zhang_pj2022}}. 

% delete old description ...

  The flare ribbons in AIA 304 $\rm \AA$ \jz{are} displayed in Figure \ref{fig:qsl_2d}(a) \jzn{and} the \jz{sigmoidal hot \jzl{channel}} in AIA 94 $\rm \AA$ \jzl{is} shown in Figure \ref{fig:qsl_2d}(b). The calculated $Q$ map at the lower layer is displayed in Figure \ref{fig:qsl_2d}(c), while \jzl{the one} along the vertical dot-dashed line in Figure \ref{fig:qsl_2d}(c) is shown in Figure \ref{fig:qsl_2d}(d). \jzl{The contours of $Q$ are overlaid on AIA 304 and 94 $\rm \AA$, respectively. }
%  A detailed analysis of the 3D magnetic topology based on CFITS can be found in \citet{Zhao_jie2016}. Briefly, the footprints of QSLs at the photosphere (Figure \ref{fig:qsl_2d}(c)) are identified to match the position and shape of the double J-shaped flare ribbons at the beginning of the flare. \jz{There are several different magnetic connectivities viewed from the $Q$ map \jz{in} Figure \ref{fig:qsl_2d}(e).} The inner centered small circle represents the core part of MFR0 with a larger twist, \jz{while} its periphery is wrapped by a relatively weakly twisted magnetic structure. In addition, a hyperbolic flux tube (HFT) configuration appears at the intersection of QSLs with an altitude about 5 Mm, indicating \jz{the possible location of} reconnection.
  \jz{\jzl{An S-shaped QSL} \jz{\jzl{is} identified in Figure \ref{fig:qsl_2d}(c)}. 
  The S-shaped QSLs generally coincide with the hot channel (L1) in AIA 94 $\rm \AA$ and its hook parts roughly match the brightenings of the flare ribbons (R1 and R2) in AIA 304 $\rm \AA$. \jzl{Such correspondence has also been derived in \citet{Zhao_jie2016} with a NLFFF-based method called Current-Field Iteration in Spherical Coordinates \citep[CFITS;][]{Gilchrist2014_CFITS}.}
  
  %, and the bright structure L1 is surrounded by the left hook. However, the peripheral part in the right corner is inharmonious compared to AIA/304 $\rm \AA$ and CFITS results. 
%an inner circle-shaped QSL, which indicates 

  In the vertical slice in Figure \ref{fig:qsl_2d}(d), \chg{three white dashed lines are displayed to provide reference for the boundaries of the flux ropes. As the QSLs obtained in the present work have many fine structures, different from the ones obtained by force-free extrapolation \citep[such as in ][]{Zhao_jie2014,Zhao_jie2016}, the flux rope boundaries are selected manually according to the rough locations with large $Q$ values. 
%  not only to the locations with large $Q$ values, but also to the general distribution of the field lines. 
  The inner circle-shaped QSLs, which indicate a strongly twisted core part of MFR1}, is wrapped by the outer QSLs.  The height of MFR1 is about 8 Mm, which is lower than \chgt{the MFR} derived by CFITS ($\sim25$ $\rm{Mm}$). A second main flux rope (MFR2) is located adjacent to MFR1 and the main body is also outlined with \chg{the} dashed line. The boundary of MFR2 is not as apparent as MFR1 in the $Q$ map, \chg{which is possibly due to the slice location that is not placed at the well-twisted region of field lines}. Moreover, there are high $Q$ regions at the bottom below 5 Mm, yet the existence of hyperbolic flux tube (HFT)\jzl{, which indicates the most probable reconnection site among the high $Q$ regions, } cannot be distinguished effectively. \jzl{It implies} that the possible reconnection may distribute in a diffuse region, \jzn{different from} the results \jzn{of} CFITS which obtains a clear HFT beneath the magnetic flux rope.}

  \subsection{Magnetic-field Configuration and QSLs}
  
% Note 2022 08 07, delete old

\begin{figure*}[ht!]
%\plotone{mag3d_04.pdf} 
%\plotone{mag3d_05.pdf} %%====================================
%\plotone{mag3d_06_5_dpi400.pdf} %%============================
%\plotone{mag3d07_dpi400_only_mhs.pdf} %%=======================
\plotone{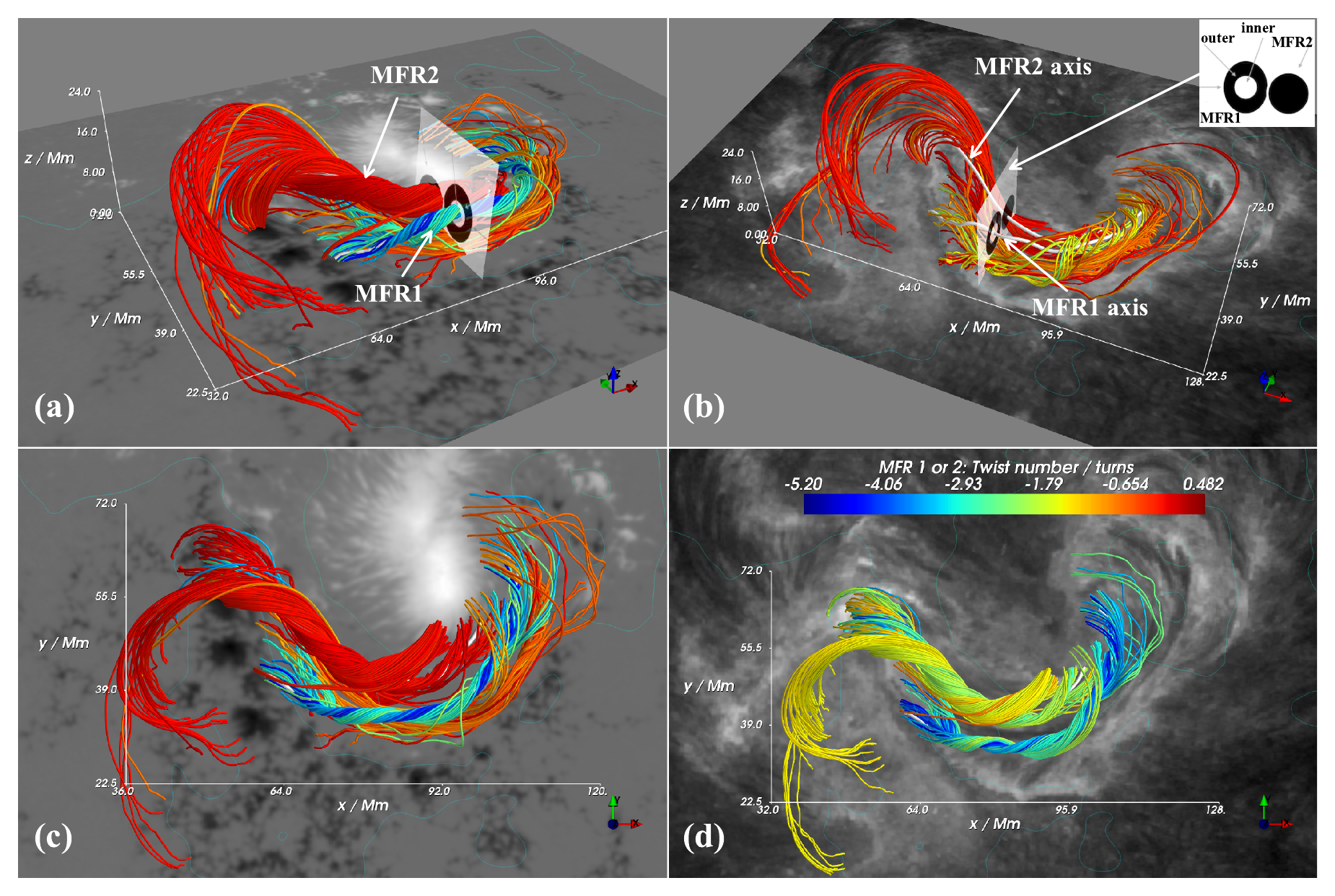} %%==============3D==========
\caption{\chgt{Magnetic connectivity and twist number of the flux rope system are displayed. All field lines are colored with the twist number and the colorbar is shown in panel (d). 
One common axis (MFR1 axis) is selected for calculating the twist number in panels (a)--(c), while two separate axes, i.e., MFR1 axis and MFR2 axis,  are adopted in panel (d) for calculating the twist number of MFR1 and MFR2, respectively. 
%The field lines in panels (a) - (c) share the color table in panel (d), while panels (e) - (g) share the color table in panel (h).  
The 2D vertical slice with black-white sketch as shown in the corner of panel (b) is inserted in panels (a) and (b) with a view angle of 3D for emphasizing the connectivity. 
}
%
% 【2022 12 05】 delete CFITS
%
\label{fig:mag3d}}
\end{figure*}

\chg{Magnetic configuration and related QSLs are explicitly revealed in Figure \ref{fig:mag3d} from 3D perspective. }%  \jz{Correspondence} between QSLs and magnetic configuration is \jz{explicitly} revealed in Figure \ref{fig:mag3d} from 3D perspective.
\chgt{As the QSLs at the vertical slice have tremendous fine structures, we have used a vertical black-white sketch, as shown in the upper right corner of panel (b), as a proxy to better display the correspondence. The sketch is obtained by extracting the main structures associated with the magnetic flux ropes as we have annotated with dashed lines in Figure \ref{fig:qsl_2d}(d). The outer part of MFR1 is displayed with black while the inner part inside is shown with white. As no fine structures are identified to be associated with MFR2, the whole flux rope region is shown with black. The annotations in the sketch are exactly the same as in Figure \ref{fig:qsl_2d}(d). The colors of the field lines in Figures \ref{fig:mag3d}(a)$-$(c) are coded with the twist number that is calculated through one common axis, i.e., MFR1 axis.} 
  %As we also calculate the twist numbers of MFR1 and MFR2 according to each axis respectively, the field lines in panel (d) are coded with two sets of twist number.
   More details of the twist number calculation is described in Section \ref{subsec:twist}.

% 【2022 12 05】 delete CFITS
%  
  
  \chgt{The \jzl{displayed} field lines reveal a complex connectivity.   The magnetic field lines in the lower altitude \jzn{(the aforementioned diffuse regions of possible reconnection from the distribution of QSLs)} are \jzl{displayed in Figure \ref{fig:mag3d}(b) individually. They are roughly along the PIL, and they show the braiding characteristics and a coupling with the axis of MFR1 to a certain extent.} } %\citep{Berger2009, Awasthi2018} 
%  The magnetic field lines in the lower altitude which are roughly along the PIL also have the braiding characteristics (panel (g))  %\citep{Berger2009, Awasthi2018} 
%  and a coupling with the axis of MFR1 to a certain extent.{\color{red}( )}
  \chgt{The core field lines of the MFR1 and MFR2 are displayed \jzl{separately} in Figure \ref{fig:mag3d}(d), \jzl{according to the QSLs as shown in the black-white sketch}.
  %\jzl{???Around the regions of MFRs indicated by the QSLs as shown in the black-white sketch}, 
  A slightly smaller twist is confirmed for MFR2, which is located in the north of the highly twisted MFR1.
}

\chgt{
  The above results again \jzn{explicitly show} that, unlike the standard single-core MFR picture appearing in other cases \citep{Zhao_jie2016, Kilpua2021, Jiang2021_nature}, the sigmoidal configuration obtained in MHS extrapolation primarily consists of two independent but slightly coupled MFRs\jzn{. The latter one, together with the ambient field lines, \jzl{corresponds} well }to the double J-shaped hot channel in AIA 94 $\rm \AA$ as shown in Figure \ref{fig:obs_and_mag}(g).}

%  
%  Delete big old

%Panels (a), (b) are images of AIA 94 and 304 $\rm{\AA}$ representing the high and low-temperature structures as well as the characteristics in upper and lower atmosphere, respectively. 

%\newpage
%\subsection{Twist number and Decay index} 
\subsection{MHD instabilities}
\label{subsec:twist}%%=================================================

%========
	The helical kink instability \citep[KI;][]{Torok2004} and torus instability \citep[TI;][]{Kliem2006}, as two basic mechanisms of ideal MHD instability, are usually invoked to be responsible for triggering and driving the MFR eruption, and are characterized by the parameters of twist number and decay index, respectively.
	
	\subsubsection{Twist Number}
	
	The KI suggests that the instability is triggered when the twist \citep{Berger2006} of the MFR exceeds a critical value $T_c$ \citep[well-known as $\sim$1.25 turns in a force-free MFR suggested by][]{Hood1981}. Based on \citet{Berger2006}, there are two typical definitions for the twist number. One measures the number of turns between two infinitesimally close magnetic field lines winding about each other and a fast code is developed by \citet{Liu_R2016} to calculate the twist in a volume. 
	The other defines the twist number as the field lines wind about a common axis. The twist density of a curve $\mathbf{y}(s)$ around the smooth common axis curve $\mathbf{x}(s)$ is defined as
%%eq for twist number  ref Guo 2010,2013,2017
    \begin{equation}  
     \frac{d \it{\Phi}}{ds} = \frac{1}{2\pi} \rm{\mathbf T}(\it s) \cdot \rm{\mathbf V}(\it s) \\
     \times \frac{d \rm{\mathbf V}(\it s)}{ds}, \label{eq:twist}
    \end{equation}
    where $s$ measures arc length from an arbitrary starting point on $\mathbf{x}(s)$, while $\mathbf{T}(s)$ is the unit tangent vector to $\mathbf{x}(s)$, and $\mathbf{V}(s)$ is the unit vector normal to $\mathbf{T}(s)$ pointing from $\mathbf{x}(s)$ to $\mathbf{y}(s)$. The final twist number $T$ is obtained by integrating Equation (\ref{eq:twist}) along the axis $\mathbf{x}(s)$.
    
	The \jz{application} of \jz{the latter} method is to find a common axis among the magnetic field lines \jz{at the first step}, and then \jz{to} calculate the twist number of each curve around the axis. \jz{The axis and computational region are determined naturally by QSLs as they could depict the main body of an MFR \citep{Guo_yang2013, Guo_yang2017_twist, Guo_JH2021}.}
	
	To better interpret the physical meaning, we adopt the latter definition to calculate the twist number with the code from \citet{Guo_yang2010, Guo_yang2013}\footnote{\url{https://github.com/njuguoyang/magnetic_modeling_codes}}. 
	%The results are shown in Figure \ref{fig:mag3d} bottom two panels (d) and (h) where the twist number for magnetic field lines are coded by colors. 
	\chgt{
 %For the results from MHS, two individual magnetic flux ropes that are weakly coupled are obtained. 
 One common axis, i.e., the axis of MFR1, has been tested for calculating the overall twist number and the results are shown in Figures \ref{fig:mag3d}(a)--\ref{fig:mag3d}(c). As the main body of MFR2 departs from the axis of MFR1, \jzl{the twist number of MFR2 is small in this manner. A} separate axis is therefore selected for MFR2 and the twist number is calculated \jzl{individually} for MFR1 and MFR2, as demonstrated in Figure \ref{fig:mag3d}(d).}

% 【2022 12 05】 delete
%	For CFITS results, MFR0 has an average twist number of about ... 
%
%
	%Meanwhile, Figure \ref{fig:mag3d} (a) - (c) has the same colormap as (d) also displaying the twist distribution of other magnetic field lines from different perspectives. 
	The \jz{highest} twist number can reach 3 turns for MFR2 and 5 turns for MFR1, \jzl{and the average values are} about 1.6 and 3 turns, respectively. 
	%among which the average value of MFR2 is about 1.6 turns ,.... Therefore, 
	The results show that, in this case, the MHS equilibrium has larger twist numbers than $T_c$ for triggering eruption. The quantitative results of twist number in the present work and other investigations are also listed in Table \ref{tab:res_compare}. \chg{As there are inner part and outer part for MFR1 as displayed in Figure 5b, the twist number is calculated referring to the same axis but for two conditions, with one for the inner part only and the other one for the overall of the flux rope including the inner part and outer part. As MFR2 has less complex distribution of the magnetic field lines, twist number has been calculated referring to MFR2 axis for the overall flux rope.}

%\chg{Note that the twist numbers of MFR1 and MFR2 are calculated according to their own axes.}  Two \jzl{conditions} are considered for MFR1 in the present work, i.e., the internal core part and the one includes both the internal and external parts \chg{(also see the dashed lines marked by 'inner' and 'outer' in Figure \ref{fig:qsl_2d}(d))}.
 %, due to the associated double-layer QSLs as shown in Figure \ref{fig:qsl_2d}(d)}. %{\color{red} what's the difference? inner/outer}

    \subsubsection{Decay Index}
    
\begin{figure*}[ht!]
\centering
\includegraphics[scale=1.1]{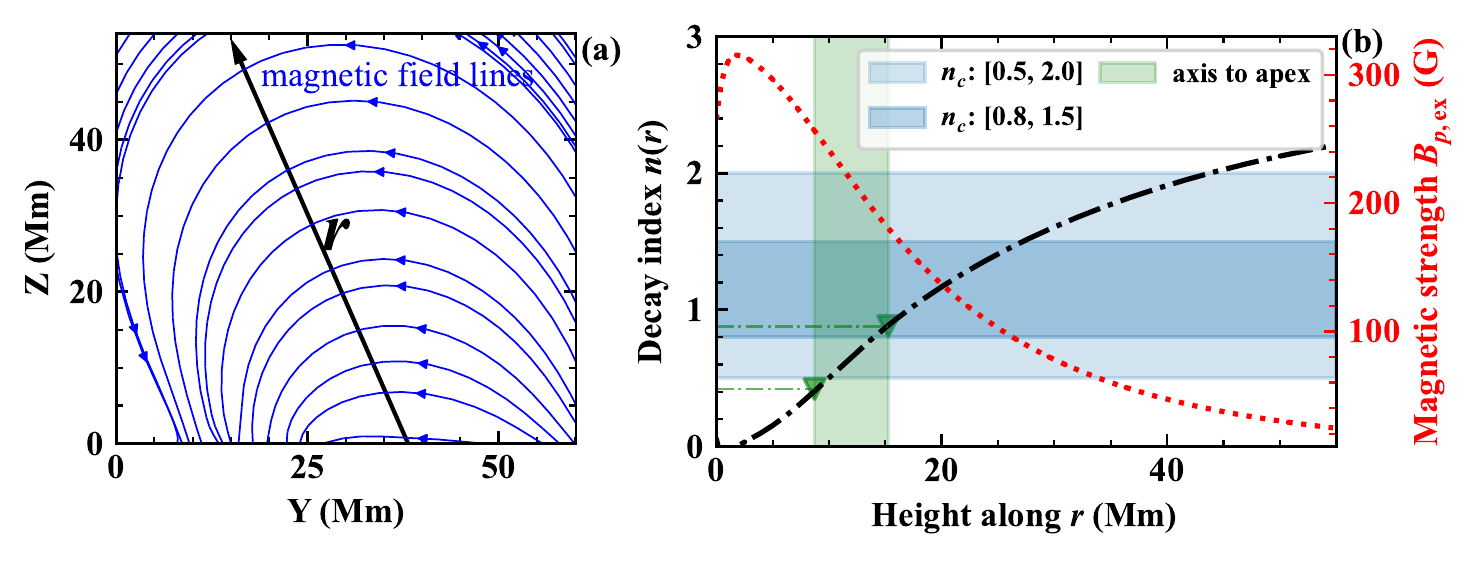}
\caption{\chgt{
Decay index. Panel (a) displays the vertical slice like those in Figure \ref{fig:qsl_2d}(d) but with field lines which tracing the projected component of the potential field in the Y-Z plane, and the oblique line with arrow depicts the direction $r$ for calculating the decay index. Panel (b) shows the profiles of the decay index $n(r)$ (black dot-dashed line) and of the strength of external poloidal magnetic field $B_{p,ex}$ (red dotted line). The blue and light blue shaded areas denote the reference range of critical decay index $n_c$ from \citet{ Zhong2021NC} and  \citet{Sun_XD2022}, respectively. The green shaded area shows the height ranges from the MFR core to the apex along the direction of $r$. The lower and upper limits of decay index are marked by $\bigtriangledown$.
}
%
% 【2022 12 05】 delete CFITS
%
\label{fig:decay_index}}
\end{figure*}    
    
    	The TI is suggested to have a significant effect on the eruption of an MFR due to the strapping force of the background field. Recently, a more relevant field \citep[$B_{p,\mathrm{ex}}$;][]{Duan2019, Zhong2021NC}, which is the component of the potential field that is perpendicular to both the axis and the erupting path of the MFR, is adopted for calculating the decay index.
%	The TI is suggested to have a significant effect on the eruption of an MFR due to the strapping force generated by the external poloidal magnetic field ($B_{p,\mathrm{ex}}$), \fy{which is the component perpendicular to both the axis and the erupting path of the MFR \citep[see][for a schematic interpretation]{Duan2019}, and usually decomposed from the potential field. }%\fy{ }
	Like in other works \citep{Bateman1978, Jiang_chaowei2014}, the decay index of $B_{p,\mathrm{ex}}$ is written as 
    \begin{equation}  
    n(r) = - \frac{d\, \mathrm{log}(B_{p,\mathrm{ex}})}{d\, \mathrm{log}(r)},\ \mathrm{or}\ 
    n(r) = -\frac{r}{B_{p,\mathrm{ex}}}\frac{d B_{p,\mathrm{ex}}}{d r}.
    \label{eq:decay_index}
   \end{equation}
   The obtained decay indices of the MHS equilibrium are shown in Figure \ref{fig:decay_index}. The black arrow $r$ in panel (a) indicates the path to calculate decay index, which roughly follows the direction of the ridge of the magnetic field lines. 
   We find that it has the decay index of about 0.4 $\sim$ 0.9, \chg{as shown in Figure \ref{fig:decay_index}(b)}. 
    \chgt{Critical range of the decay index has been found within 0.8 $\sim$ 1.5 in \citet{Zhong2021NC} and within 0.5 $\sim$ 2.0 in \citet{Sun_XD2022}. 
    According to such criteria, the MHS model in the present work is able to trigger torus instability. Nevertheless, %comparing to the decay index from CFITS,% 
    the relative small index of MHS might be related with the low-lying flux rope, where the strapping field above could still be relatively strong.}

\subsection{Magnetic Energy and Helicity} \label{subsec:e_and_hm}%%==============
\begin{figure*}[ht!]
%\plotone{en_hm2d_v3_.pdf}    %%======================================
%\plotone{en_hm2d_v9_just_energy.pdf}    %%============================
%\plotone{en_hm2d_v8.pdf}    %%========================================
%\plotone{en_hm3d_render4.pdf}    %%==================================
%\plotone{en_hm3d_render5.pdf}    %%==================================
\centering
\includegraphics[scale=0.785]{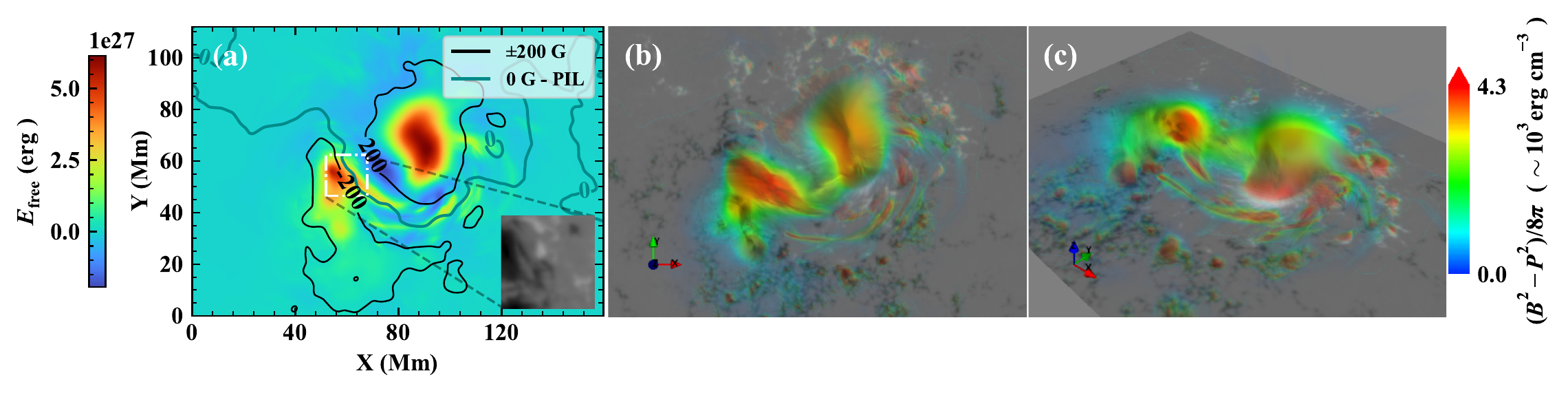}
%\includegraphics[scale=0.72]{en_mhs_cfits2d_3d.pdf} %2022 12 05 
%\includegraphics[scale=0.82]{en_hm3d_render9_dpi300.pdf}
%\includegraphics[scale=0.82]{mag3d_render_9_dpi300_just_energy.pdf}
%\plotone{mag3d_render_9_dpi300_just_energy.pdf}
%\includegraphics[scale=0.82]{en_hm3d_render6_dpi400.pdf}
\caption{ \chgt{
Distributions of free energy in 2D and 3D. The 2D map in panel (a) displays the integrals along $Z$ direction with several layers covering the primary structure. The contours in cyan and dark show the $B_z$ component from the potential field with levels of 0 and ±200 G as in Figure \ref{fig:qsl_2d}. The white dot-dashed box denotes the site of shearing flow and magnetic field cancellation suggested by \citet{Cheng_xin2015}.
The 3D maps in the right two panels show the volume rendering of $(B^2-P^2)/8\pi$ \jzl{with bird view and side view, respectively}}.
%}
%
% 【2022 12 05】 delete CFITS
%
\label{fig:en_hm2d}}
\end{figure*}

%--old before 2022 10 20
%delete

Magnetic free energy and magnetic helicity, which are frequently investigated in previous studies, are also explored in the present work. Magnetic free energy is believed to be the upper limit of the energy release, while magnetic helicity essentially manifests the complexity of the magnetic field \citep{Berger1984}. 
%\chg{As both calculations commonly require the 3D potential field as reference, the Laplace's equation is solved numerically with Neumann boundary condition.
%} 
%2023 03 21 note
%\chg{As both calculations commonly require the 3D potential field as reference, the Laplace's equation ...
%} 

% and spherical geometry (see Appendix \ref{sec:green_fun}). 
%The details of the calculations are described as follows.

\subsubsection{Magnetic Energy}

%The distribution of magnetic energy and helicity in both 2D and 3D as well as quantitatively numerical results will be displayed below as they are two important aspects to investigate solar eruptions. 
        
The energy stored in magnetic field that can power eruptions is the free energy, which is estimated as:
\begin{equation}  
 E_{\rm{free}} = \int_V{\frac{B^2}{8\pi}}dV-\int_V{\frac{P^2}{8\pi}}dV = \int_V{\frac{B^2-P^2}{8\pi}}dV,
\end{equation}
%% E_{\rm{free}} = E_{\rm{B}}-E_{\rm{P}} = \int_V{\frac{B^2}{8\pi}}-\int_V{\frac{P^2}{8\pi}}
where $B$ and $P$ are the strength of non-potential field and reference potential field respectively. \chg{The latter one is obtained based on the Green's function method \citep{Chiu1977, Sakurai1982, Wiegelmann2005}, in order for the comparison of free energy in Section \ref{sec:res_comp}.} 
The integrand denotes the energy density distributed over the computational domain $V$. 
\jz{By integrating energy density along $Z$ direction \chg{and multiplied by the surface element $dS=dxdy$ \citep{Jiang_chaowei2014}}, a 2D distribution of free energy $E_{\rm free}(x,y)$ is obtained} as shown in Figure \ref{fig:en_hm2d}(a). %and a similar demonstration was adopted by \cite{Mackay2011} and \cite{Jiang_chaowei2012}. 
%\fy{It should be noted that the free energy in each part of the distribution cannot be guaranteed to be nonnegative physically, but it is meaningful for the whole volume integral as positive \citep[e.g.,][]{Mackay2011, Jiang_chaowei2014}}. 
%The results indicate that the free energy of CFITS is ,
\chgt{The result indicates that the free energy of MHS has high values} distributed along MFR1 and MFR2 (also see Figure  \ref{fig:en_hm2d} \jzl{(b-c)} for 3D renderings). 
It is worth noting that the MHS model has conspicuous free energy in the region that is outlined with a white box in Figure \ref{fig:en_hm2d}(a), where \citet{Cheng_xin2015} investigated the shearing flow and magnetic cancellation at the same region. We also notice the high values in the sunspot region. 
As discussed by \citet{Borrero2011} and \citet{Tiwari2012}, sunspots tend to be non-potential which implies that the existence of free energy, and the consideration of plasma effect in MHS model may also lead to free energy. 
% delete neg-energy dis

\subsubsection{Magnetic Helicity} \label{subsec:mag_h}
Magnetic helicity, in its original definition, is the volume integral of the scalar product between vector potential and vector magnetic field. \jz{It} measures the linkage of a bundle of magnetic field lines, and is related to the topological concepts such as Gauss linking number, \jz{and} the usually used twist number and writhing number \citep{Berger1984, Moffatt1992, Berger2006}. 
%The original form is suitable for closed magnetic configuration, we therefore use 
\jz{For solar active region, relative magnetic helicity \chg{as a conserved quantity} is given by \citet{Berger1984} and \citet{Finn1985} \chg{for the domain of volume $V$:}}
%% Helicity define,  as the gauge-invariant formulation:
\begin{equation}  
 \Hm = \int_{V}{ (\mathbf A+ \Ap) \cdot (\mathbf B - \mathbf P) }\, dV, \label{eq:hm}
\end{equation}
where $\mathbf A$ and $\Ap$ are the vector potentials of $\mathbf{B}$ and $\mathbf{P}$. \chg{Potential field $\mathbf{P}$ is usually adopted as the reference field. The boundary condition $(\mathbf {\hat{n}} \cdot \mathbf{B}) \mid_{\partial V} = (\mathbf {\hat{n}} \cdot \mathbf{P}) \mid_{\partial V}$ is adopted here to keep $\Hm$ gauge-invariant}. This form of helicity calculation \chg{used within a considered volume} is also known as the finite volume method, \chg{which enables a direct comparison of $\Hm$ computed from different magnetic fields \citep{Valori2012}}. More methods of helicity estimation have been proposed, such as twist-number (TN), helicity-flux integration (FI), connectivity-based (CB), see \citet{Valori2016}, \citet{Guo_yang2017} and \cite{Thalmann2021} for an overview. 
%
%delete lagrely

%\subsubsection{Current-carrying Helicity and Twist Helicity}

As shown in previous works \citep{Berger1999, Berger2003, Guo_yang2017_twist}, the helicity in a finite volume can be decomposed into two components, with one component representing the contribution from the magnetic field that carries local currents and the other component standing for the contribution from the potential field and magnetic field generated by local currents.
The \jz{former component is called current-carrying helicity as $H_{\rm J}$ in the present work and} is calculated as follows:
\begin{equation}
H_{\rm J} = \int_V{(\mathbf{A}-\mathbf{A}_{p})\cdot(\mathbf{B}-\mathbf{P})}\, dV . \label{eq:H_J}
\end{equation}

Solar eruptions often show highly twisted magnetic flux rope, which initiated a twist number method for the helicity calculation \jz{\citep{Guo_yang2010,Guo_yang2013,Guo_yang2017_twist}}. \jz{By neglecting the contribution from writhe, and also the one from the mutual helicity between magnetic flux rope and ambient field,} the twist helicity is calculated as follows:
\begin{equation}
 H_{\rm{twist}}\approx T\Phi^2 ,
\end{equation}
where $\Phi=\iint{B_x\, dydz}$ is the axial magnetic flux. 

\chgt{The relative helicity, current-carrying helicity and twist helicity are then obtained \jzl{and are} shown in Table 1.
It is found that $|H_{\rm{twist}}|$ of the MHS model is smaller than $|H_{\rm{J}}|$, which is probably due to the fact that the local \jz{current} is not completely limited to the volume \jz{occupied by} the MFR. 
%{.} 
In addition, the MHS model} has two coupled MFRs and braiding with the surrounding magnetic field, thus the mutual helicity is underestimated in the computation. 
%Meanwhile, $H_{\rm twist}$ is also affected by $\Phi^2$ because of the smaller magnetic flux.

%=========ratio=========
%{\color{magenta} 
The role of magnetic helicity for indicating the solar eruptions has been extensively discussed \citep{Toriumi2022}. MHD simulations 
\citep[e.g.,][]{Pariat2017,  Zuccarello2018, Linan2018} suggest that helicity ratio,  i.e., $|H_{\rm J}|/|H_{\rm m}|$ is a promising proxy of solar eruptivity.  
In observations,  statistics \citep{Gupta2021} and case studies \citep{James2018, Moraitis2019, Thalmann2019a} using force-free modeling also show that $|H_{\rm J}|/|H_{\rm m}|$ has advantages in predicting the eruptive potential of a flaring AR.
\fyt{  
It is mainly characterized by the higher $|H_{\rm J}|/|H_{\rm m}|$ values \citep[e.g., \textgreater 0.1 in][]{Gupta2021} for CME-productive active regions, and the reduction of $|H_{\rm J}|/|H_{\rm m}|$ during eruptions.} 
\chg{In the present work, the helicity ratio given by MHS is about 0.06.} 
By inspecting the MHS data, 
we find that there are some positive values in the volume of integrand from equation (\ref{eq:H_J}), 
although the integral helicity of the entire computational volume is negative. \chg{The absolute ratio of the positive integral to the negative one of $H_{\rm J}$ is about 0.5}, which naturally leads to a small value of current-carrying helicity integration as the two components are not treated separately. Such difference is inherently dependent on the \jzl{model that is adopted for reconstructing the magnetic configuration.}
%two methods which reconstructing different magnetic configurations.
%\chg{By inspecting the MHS result, we find that there are some positive values of $H_{\rm J}$ distributed at the right end of MFR1 and its peripheral regions (see Figure \ref{fig:mag3d}(f) for reference), and the ratio of the positive part to the negative part of $H_{\rm J}$ is about 0.56 for the integral value of the whole computational region, which naturally leads to a small $|H_{\rm J}| / |H_{\rm m}|$.
Nevertheless, there are exceptions of the statistical threshold \citep{Thalmann2021, Gupta2021} and whether there is a universal criterion for extrapolations with different assumptions, \chgt{e.g., the force-free and non-force-free ones}, requires further examinations with more samples \citep{Moraitis2019, Toriumi2022}.
%}

\subsubsection{Comparison with previous results}\label{sec:res_comp}
  AR 12158 has been extensively studied \jz{in various aspects and with various extrapolation methods}, here we present some comparisons between our results and others in Table 1. 
%  and their associations with previous results. 
%\jz{Considering different extrapolation methods, a quantitative comparison is listed in Table \ref{tab:res_compare} }.
%in the aspect of magnetic energy and magnetic helicity.} 
\chgt{\jzl{\citet{Duan2017} compared different extrapolations with NLFFF code \citep[CESE-MHD-NLFFF, by][]{Jiang2013} and non-force-free field code \citep[NFFF, by][]{Hu2008, Hu2010}. Comparing to NLFFF2 and NFFF, MHS obtains free energy and relative helicity in the same order of magnitude. In the aspect of magnetic configuration, MHS obtains two twisted flux ropes while the other methods obtained sheared arcades or one coherent flux rope with medium twist number.}
%Table \ref{tab:res_compare} also shows results from \citet{Duan2017} who compared different extrapolations with NLFFF code \citep[CESE-MHD-NLFFF, by][]{Jiang2013} and non-force-free field code \citep[NFFF, by][]{Hu2008, Hu2010}. 
%MHS obtains free energy and relative helicity in the same order of magnitude, comparing to NLFFF2 and NFFF.  
\jzl{The above results demonstrate that,} although the plasma forces bend the magnetic field lines to form a non-force-free configuration in MHS method,  the non-force-freeness does not necessarily bring in more energy or helicity.
%, and the occupation of the obtained flux rope might be important for the integral results of the free energy and relative helicity.
}

% 【2022 12 05】 delete high old
%\chg{Solution obtained with the CFITS has ...

%2022 10 08 back 
%Overall, the above results demonstrate that ...

%==copy ori delete

%%%========================================================
%%====v2=======【2022 0311】 old

%==================Table===========================

%%========================================================
%====v3=======【2022 0828】+ re-computation
\begin{deluxetable*}{cccccccccc}
\tablenum{1}
\tablecaption{\jz{Magnetic} quantities \jz{derived from different methods} \label{tab:res_compare}}   %% 
\tablewidth{0pt}

%%
%%\tablehead{
%%Model & $\Phi$ & $E_{\rm{tot}}$ & $E_{\rm{pot}}$ & $E_{\rm{free}}$ & %%$E_{\rm{free}} / E_{\rm{pot}} $ & $H_{\rm{m}}$ & $H_{\rm{J}}$  & %%$H_{\rm{twist}}$ & $\overline{|T_{\rm{w}}|}$
%%}
%%

\tablehead{
\colhead{Model} & \colhead{$E_{\rm{tot}}$} & \colhead{$E_{\rm{pot}}$} & \colhead{$E_{\rm{free}}$} & \colhead{$E_{\rm{free}} / E_{\rm{pot}} $} & \colhead{$H_{\rm{m}}$} & \colhead{$H_{\rm{J}}$} & \colhead{$H_{\rm{twist}}$} & \colhead{$\overline{|T_{\rm{w}}|}\ $} \\
\colhead{} & \colhead{($10^{32}\rm{erg}$)} & \colhead{($10^{32}\rm{erg}$)} & \colhead{($10^{32}\rm{erg}$)} & \colhead{} & \colhead{($10^{43}\rm{Mx}^2$)} & \colhead{($10^{43}\rm{Mx}^2$)} & \colhead{($10^{43}\rm{Mx}^2$)} & \colhead{(Turn)}
}

%%\decimalcolnumbers  %% 
\startdata
MHS   & 11.15 & 9.90   & 1.25 & 12.7$\%$ & \chg{-1.60} & \chg{-0.09}  & -0.007±0.004 & MFR1, 1.39±0.92\chg{$^{a}$}, 3.06±0.82\chg{$^{b}$} \\
      &       &        &      &          &       &         &   & MFR2, 1.59±0.40\chg{$^{a}$}, $\qquad \qquad \quad $ \\
%\\
NLFFF1 & 11.5 & 11.7 & -0.18 & -1.5$\%$ & -1.60 & - & - & $\leq\ $1.0 \\
NLFFF2 & 11.1 & 10.0 & 1.10  & 10.9$\%$ & -2.03 & - & - & $\leq\ $1.0 \\
NFFF   & 14.8 & 11.8 & 3.02  & 25.6$\%$ & -2.36 & - & - & $\leq\ $1.0 \\
CFITS & - & -   & - & - & - & -   & - & 1 $\sim$ 2 \\
%TMFM & - & -   & - & - & - & -   & - & $\textgreater\ $1.0 \\
\enddata

\tablecomments{NLFFF1, NLFFF2, and NFFF are results from \citet{Duan2017}, and their twist number are computed by the first definition. % The real computing region not the one, some difference, but including the core region.
The NLFFF1 and NLFFF2 use unpreprocessed and preprocessed vector magnetograms as boundaries, respectively. The negative free energy in NLFFF1 arises probably due to inconsistency between the unpreprocessed magnetogram and the force-free approximation.
\chgt{CFITS \jzl{represents the result from \citet{Zhao_jie2016} with an approximate twist number}}. 
%\chg{The 'inner' is indicated only for MFR1 as shown in Figure \ref{fig:qsl_2d}(d).} 
}  %%脚注
\tablenotetext{a}{\chg{ Twist number for the overall flux rope.}}\label{note:a}
\tablenotetext{b}{\chg{ Twist number for the inner core part of MFR1.}}\label{note:b}
\end{deluxetable*}

% 【2022 12 05】 delete CFITS
%%====v3=======【2022 0828】 re-computation
%

\vspace{-3em}
\section{Summary and Discussion} 
%\section{Discussion and Conclusion}
\label{sec:discussion}%%============== =================
%Demo: Using the vector magnetic field data of SDO/HMI as the boundary condition, we extrapolate the force-field (MHS) and force-free field (CFITS), and compare the effects of the two methods on magnetic field reconstruction.
%\begin{itemize}
%\item[•] conclusions of major results 
%\end{itemize}
%\begin{itemize}
%\item[•] calculation of Decay index ?
%\end{itemize}

%\begin{itemize}
%\item[•] discuss: Braid magnetic field structure of MHS\\
%Theory and observation, ex. 
%\citet{Berger2009}, \citet{Awasthi2018}, etc
%\end{itemize}

%\begin{itemize}
%\item[•] discuss: High twist of MHS\\
%\citet{low1992}, \citet{Guo_JH2021}, \\
%sunspot rotation: \citet{Brown2021, Vemareddy2016}.  \citet{Duan2021} another AR $\sim$6 turns, 
%\end{itemize}

%%============================================================================
%%================================== ================================
\subsection{Summary} \label{subsec:Summary} %%??
%Begin:
Magnetic field in the solar atmosphere, which is responsible for solar eruptions, interacts with plasma in nature, especially at the lower atmosphere such as the photosphere and lower chromosphere where plasma $\beta$ is non-negligible. To construct a 3D magnetic field above the photosphere, which could imitate the observations in more details, it is necessary to consider the contribution of plasma effect. 
%, and the gravity and pressure gradient forces should be taken into account in the first step. 
In this paper, \chgt{we reconstruct the magnetic field of AR 12158 by using the MHS extrapolation method, \jzl{evaluate its performance in reproducing the observations in multiple wavelengths ranging from chromosphere to the corona.  
We also discuss the} magnetic topology and connectivity, \chg{decay index and twist number}, and the magnetic energy and relative helicity.
%Additionally, the performance of \jz{MHS model} for reproducing the observations in SDO/AIA EUV passbands, BBSO/GST He I 10830 $\rm{\AA}$ and IRIS/SJI 2796 and 1400 $\rm{\AA}$ are evaluated.
}

%2022 12 05 note
%In this paper, \chgt{we reconstruct the magnetic field of AR 12158 by using the MHS extrapolation method and compare it \jz{mainly} with the force-free field (CFITS method) from \citet{Zhao_jie2016}, in the aspects of magnetic topology and connectivity, MHD instabilities, magnetic energy and relative helicity.} Additionally, the performance of \jz{MHS model} for reproducing the observations in AIA EUV passbands are evaluated.

%	It is well known that the interaction between. delete CFITS

The EUV observations before eruption imply that AR 12158 contains abundant atmospheric structures, such as the S-shaped structure composed of two separated J-shaped \jz{segments} in AIA 94 $\rm{\AA}$, the dark stripes in AIA 304 $\rm{\AA}$, and the twisted \jzl{braiding} structure in multi-wavelengths. \jzl{The high resolution observations in IR and UV also show the multi-temperature fibrils. All the above mentioned fine structures, \jzl{especially the braiding structures, are in good agreement with the reconstruction results of the MHS model. The braiding required by Parker-type nanoflare scenarios prefers such a non-force-free environment \citep{Aschwanden2019} like in
the chromosphere and transition region, and it could be difficult to reproduce from the methods with force-free assumption.}
During the eruption, magnetic topology from the MHS model is co-spatial with the flare ribbons as well as the hot channel, like the one obtained by the NLFFF extrapolation \citep{Zhao_jie2016}.} 

\chgt{The MHS model distinguishes two separate but weakly coupled \jz{flux} rope systems (MFR1 and MFR2). MFR1 has an average twist of about 3 turns and a maximum one of about 5 turns for the core part, while MFR2 has a twist number of about 1 $\sim$ 2 turns which is close to the result in \citet{Zhao_jie2016} and \citet{Kilpua2021}.} 
%Sheared arcades have also been obtained for this active region by some other extrapolation methods \citep[e.g.][]{Vemareddy2016, Duan2017, He_wen2022}. 
%\jz{For most extrapolation results, the sigmoid structure from the observation has a rough counterpart in the magnetic field. Among these extrapolations,} the magnetic flux rope MFR2 of MHS is more likely in consistent with the hot channel suggested by \citet{Cheng_xin2015}. The mentioned site of shearing flow and cancellation corresponds to the location where the free energy from MHS are most concentrated.%, while no explicit correspondence can be found \jz{from other methods.}
%\jzl{According to the in-situ measurement of a twist number in the range of 1.0 $\sim$ 2.4 per astronomical unit \citep{He_wen2022}, the magnetic field obtained by MHS seems to be consistent with the observation.}
%
\jzl{For} the overall asset, such as the total amount of free energy and relative helicity, the MHS method \jz{as well as other NLFFF and NFFF methods}, derive values on the same order of magnitude, i.e., $\sim10^{32}$ erg and $\sim10^{43}\ \rm{Mx^2}$ respectively.  

\jz{
\jzl{This work shows} that although the estimated total amount of energy and helicity are on the same order for different extrapolation methods, there are differences in the specific magnetic configuration, thus leading to different understanding of the eruption. 
\chg{The force-free field models compared in this work enable a general description of the active region magnetic field . Detailed magnetic structures, which are crucial for understanding the
evolution of flux rope as well as the trigger of eruption at the initial phase, have been acquired from the MHS model with consideration of the plasma.  Beyond the case study, comparisons need to carry out with more samples and also with other extrapolation methods in future to validate the advantage of the non-force-free method in a more general sense.}
}  
%

 %% further about twist=====================================
 \subsection{Further Discussion} \label{subsec:FD} %%??
	High twist number is obtained from the MHS method while only sheared arcades or weakly twisted flux rope have been reproduced from other NLFFF methods. The validation of such high twist and the associated MHD instability are further discussed in this subsection.
	
	\subsubsection{High Twist Number}
%	Based on the methodological tests, \citet{low1985,low1992} \jz{proposed an} analytical MHS equation for solar active region.
%
%\chg{By applying an analytical MHS solution \citep{low1991} with two current systems (one following along and the other across the magnetic field) to a magnetogram, \cite{low1992}} found that the magnetic field lines become more complex with elongated and multi-helical features \chg{due to the cross-field current associated with plasma effect.}
By applying an analytical MHS solution \citep{low1991} to a magnetogram with two current systems (one following along and the other across the magnetic field), \cite{low1992} found that the magnetic field lines become more complex with elongated and multi-helical features due to the cross-field current associated with plasma effect. 
Although the features are observed in the analytical model, they provide some instructive glimpse of highly-twisted magnetic flux interacted with plasma in real atmosphere.
%It in principle demonstrates the possibility of high twist number by considering plasma when inferring the 3D magnetic field from photospheric magnetic field.
%
%when considering both field-aligned and cross-field electric currents in their model. It in principle demonstrates the effect of considering plasma \jz{when inferring the 3D magnetic field from photospheric magnetic field}. 
 %\citet{Zhu_XS_2019} tested the MHS approach with radiative MHD simulation. They found that the twisted magnetic structure is well reconstructed by the MHS model, while the NLFFF \citep{Wiegelmann2004, Wiegelmann2006} only show sheared lines.

    In the aspect of observations, the rotation of sunspot may contribute to the high twist of magnetic field lines. Such contributions have been identified in \citet{Vemareddy2016} and \citet{Brown2021} for AR 12158, and in AR 11943/11944 \citep{Duan2021} for the flux rope structure. Twist number as high as 6 turns has been obtained in the latter case. \jz{Interplanetary magnetic cloud with high twist number is frequently observed at L1 point \citep{Wang_YM2016}, indicating the eruptive counterpart near the sun may be highly twisted originally.}
    %\fyt{ In addition, the time-dependent data-driven simulation and in-situ measurements also show high twist number of the MFR structure from AR 12158 \citep{Kilpua2021}}.

    \subsubsection{MHD instability}
    MFR with large twist number is considered to be explosive due to kink instability. Both the theories and observations demonstrate that the critical twist number has a large range and is influenced by such as the external magnetic field, the plasma flow, and plasma $\beta$ \citep[see discussion in][]{Guo_JH2021}. 
    %Specifically, \citet{Srivastava2010} reported an eruption with twist number of 6 turns.

    The maximum twist number ($|T|_{\rm{max}}$) was found to be sensitive during flares \citep{Liu_R2016}. Such parameter was employed by \citet{Duan2019} as 
    \jzl{a controlling parameter for investigating the kink instability of the pre-flare MFRs in a statistical manner.}
    %the \chg{kink instability} controlling parameter of the pre-flare MFRs in a statistical study. 
    Their results, \jzl{which are based on the force-free assumption,} demonstrate that flare events with $|T|_{\rm{max}}$ exceeding its critical value ($|T|_{\rm{c}}$) have a high possibility to \jz{erupt} successfully. Lower limit and average values of $|T|_{\rm{c}}$ from statistics are suggested to be 2 and $2.83\pm1.31$ turns respectively. 
    %The obtained $|T|_{\rm{max}}$ values of MFR1 ($\sim$5 turns) and MFR2 ($\sim$ 3 turns) thus indicate high possibility of KI. 

	\jz{Meanwhile,} previous works \citep{Dungey1954, Hood1979, Bennett1999, Baty2001} estimated the critical value, \jzl{$T_c = \frac{\omega_c L}{2 \pi a}$}, of a uniform twisted MFR with axial length $L$ and minor radius $a$, suggesting a higher value for thinner MFR. 
    \jz{Considering the non-force-free state in the present work,} the parameter $\omega_c$ {, which depends on detailed configuration of an MFR,} is \jz{selected to be around} 2 \citep{Dungey1954, Bennett1999, Wang_YM2016}. With the aspect ratio $L/a$ being estimated as 18 for MFR1, we derive a critical twist of about 5.7 turns which is greater than the $|T|_{\rm{max}}\approx 5$ turns of MFR1. \jzl{Under this estimation, the stability of MFR1 is consistent with the half eruption that discussed in \citet{shen_jinhua2022} where they show the erupting of the left J-shaped sigmoid that corresponds to MFR2 in the present work.}\\
    
%    \jzl{support？}
%    \subsubsection{Possible Braiding}
%     The braiding mechanism of magnetic field lines provides a scheme for \jz{energy} dissipation and coronal heating in the parker-type nanoflaring scenarios \citep{Parker1988, Berger2009, Aschwanden2019, Pontin2020}. \citet{Cirtain2013Nature} \jz{reported} the braided fine structure from Hi-C high-resolution observations. \citet{Awasthi2018} also revealed the pre-eruptive reconnection in a \jz{braided} multi-flux-rope system. In addition, \citet{Aschwanden2019} \jz{suggested} that the braiding required by parker-type scenarios prefers a non-force-free environments, such as in the chromosphere and transition region.
%     In the present work, the MHS technique \jz{reproduces} magnetic correspondence of the distinct braiding features, which is not produced by the force-free field method. Therefore, the comparison between MHS and \chgt{observations} indicates that the MHS \jzn{method} might be conducive to reproduce \jz{fine structures such as} the braided characteristics.\\

%This work is supported by National Natural Science Foundation of China, Grant No. 12233012
%===========================================================================
%========Thanks========
Acknowledgements. 
%\jzn{We thank the anonymous referee and scientific editor Dr. Manolis K. Georgoulis for their valuable suggestions.}
We thank Dr. Yingna Su for an internal review of this paper. We also thank the open-source tools 
(\href{https://docs.enthought.com/mayavi/mayavi/}{Mayavi} and \href{https://www.paraview.org/}{ParaView}) for 3D scientific data visualization. 
Data from observations are provided by NASA/SDO and the HMI and AIA science teams, \jzl{by the IRIS and BBSO scinece teams. IRIS is a NASA small explorer mission developed and operated by LMSAL with mission operations executed at NASA ARC and major contributions to downlink communications funded by ESA and the Norwegian Space Center. 
BBSO operation is supported by NJIT and US NSF AGS-1821294 grant. GST operation is partly supported by the Korea Astronomy and Space Science Institute, the Seoul National University, and the Key Laboratory of Solar Activities of Chinese Academy of Sciences (CAS) and the Operation, Maintenance and Upgrading Fund of CAS for Astronomical Telescopes and Facility Instruments.
F.Y., J.Z., and Y.S. acknowledge the support by 2022YFF0503001 and National Natural Science Foundation of China (grant Nos. 11820101002). J.Z. acknowledges the supports by National Natural Science Foundation of China, Grant No. 12233012, 11503089, U1731241 and Chinese Academy of Science Strategic Pioneer Program on Space Science, Grant No. XDA15052200, XDA15320103, XDA15320301, and mobility program (M-0068) of the Sino-German Science Center. 
X.Z. acknowledges financial support by National Key R\&D Program of China (2021YFA1600503), mobility program (M-0068) of the Sino-German Science Center, and NSFC grant 11790301. 
Y.G. was supported by NSFC (11773016, 11961131002, and 11533005) and 2020YFC2201201. This work is supported by Open Research Program No. KLSA202112 of CAS Key Laboratory of Solar Activity. 
\chg{This research is supported by the International Space Science Institute (ISSI) in Bern and the ISSI-BJ in Beijing, through ISSI International Team project 568 and ISSI-BJ International Team project 55 (Magnetohydrostatic Modeling of the Solar Atmosphere with New Datasets).}
}

%Acknowledgements. \jzn{We thank the anonymous referee and scientific editor Dr. Manolis K. Georgoulis for their valuable suggestions.
%We thank Dr. Yingna Su for an internal review of this paper.} 
%Data from observations are provided by NASA/SDO and the HMI and AIA science teams. We also thank the open-source tools 
%(\href{https://docs.enthought.com/mayavi/mayavi/}{Mayavi} and \href{https://www.paraview.org/}{ParaView}) for 3D scientific data visualization. 
%This work is supported by Open Research Program No. KLSA202112 of CAS Key Laboratory of Solar Activity. 
%J.Z. acknowledges the supports by National Natural Science Foundation of China, Grant No. 12233012, 11503089, U1731241 and Chinese Academy of Science Strategic Pioneer Program on Space Science, Grant No. XDA15052200, XDA15320103, XDA15320301. 
%F.Y., J.Z., and Y.S. also acknowledge the National Natural Science Foundation of China (grant Nos. 11820101002). X.S. acknowledges financial support by National Key R\&D Program of China (2021YFA1600503), NSFC grant 11790301, and mobility program (M-0068) of the Sino-German Science Center. Y.G. was supported by NSFC (11773016, 11961131002, and 11533005) and 2020YFC2201201.

%===========================================================================
%===========================================================================
%==============================【公式附录】===================================
%===========================================================================
%===========================================================================
%
\appendix 
\section{Supplement on magnetic helicity}
\label{sec:mag_helicity_add}

\chg{
The potential field solved with either one bottom boundary or six boundaries can be conveniently used for the calculation of relative magnetic helicity under the original DeVore's method \citep{DeVore2000_2, DeVore2000} or the DeVore-gauge based finite volume (FV) method \citep[e.g., using DeVore$\_$GV gauge in][]{Valori2012}, respectively. 
For the latter one, there are many ways to solve the 3D Laplace problem numerically as the first step, such as the Helmholtz solver in the Intel$^\circledR$ Mathematical Kernel Library (MKL), the HW3CRT routine in the FISHPACK library \citep{Swarztrauber1979}, or the fast Fourier transform (FFT) based method as described in \citet{Valori2016}.}

\chg{
It should be noted that the DeVore's method is defined in semi-infinite space $(z \geq 0)$. In practice, the volume integrals is inevitably limited to a finite volume, so the choice of integral volume destroys the original gauge invariance. 
The FV method is defined in a finite volume, which provides a way to make the helicity values from different magnetic fields comparable. Therefore, the FV method is recommended and widely used in the comparative or statistical study related to relative magnetic helicity \citep[e.g., ][]{Gupta2021} in recent years, and is also adopted in this work. 
}

% 2023 04 03，注释掉Devore的方法公式

\chg{
According to the FV method from \cite{Valori2012}, we compute $\mathbf{A}$ and $\Ap$ under DeVore$\_$GV gauge in a volume $V = [x_1, x_2] \times [y_1, y_2] \times [z_1, z_2]$ as follows: 
\begin{eqnarray} %%  
 \mathbf A &=& \mathbf b + \mathbf {\hat{z}} \times \int_z^{z_2}{\mathbf B}\, dz',\label{eq: A1} \\
 \Ap &=& \mathbf b_p + \mathbf {\hat{z}} \times \int_z^{z_2}{\mathbf P}\, dz',
\end{eqnarray}
where $\mathbf b = \mathbf b_p \equiv \overline{\mathbf b}$ is a particular solution specified and proved by \cite{Valori2012} as follows:
\begin{eqnarray} %% B  
  \overline{b}_x &=& -\frac{1}{2} \int_{y_1}^{y}{B_z(x, y', z_2)}\, dy', \\
  \overline{b}_y &=&\ \ \frac{1}{2}\int_{x_1}^{x}{B_z(x', y, z_2)}\, dx'.\label{eq: A1_}
\end{eqnarray}
}

\chg{
%\eqref{eq: A0} -- \eqref{eq: Ap1} and
Equations \eqref{eq: A1} -- \eqref{eq: A1_} is a performable scheme for deriving vector potential before computing helicity with FV method. The method} as carried out above estimates a global quantity of magnetic helicity. Moreover, one could refer to the relative field line helicity \citep[e.g., ][]{Yeates2018_dec, Moraitis2019_apr, Moraitis2021_may}, which is developing vigorously in recent years, for a distribution investigation.

%2023 02 

%2022 12 05 删除球坐标的势场，无球坐标需求

%\section{Using Chinese, Japanese, and Korean characters}

%Authors have the option to include names in Chinese, Japanese, or Korean (CJK) 
%characters in addition to the English name. The names will be displayed 
%in parentheses after the English name. The way to do this in AASTeX is to 
%use the CJK package available at \url{https://ctan.org/pkg/cjk?lang=en}.
%Further details on how to implement this and solutions for common problems,
%please go to \url{https://journals.aas.org/nonroman/}.


%%===========================================
\bibliographystyle{aasjournal}



\end{CJK*}
\end{document}